\documentclass[useAMS,usenatbib]{mnras}

\usepackage{graphicx,natbib,url,twoopt}

\usepackage{amsmath}
\usepackage{amssymb}

\bibpunct{(}{)}{;}{a}{}{,}    
\makeatletter
 \newcommandtwoopt{\citeads}[3][][]{\href{http://adsabs.harvard.edu/abs/#3}%
   {\def\hyper@linkstart##1##2{}%
    \let\hyper@linkend\@empty\citealp[#1][#2]{#3}}}    
 \newcommandtwoopt{\citepads}[3][][]{\href{http://adsabs.harvard.edu/abs/#3}%
   {\def\hyper@linkstart##1##2{}%
    \let\hyper@linkend\@empty\citep[#1][#2]{#3}}}      
 \newcommandtwoopt{\citetads}[3][][]{\href{http://adsabs.harvard.edu/abs/#3}%
   {\def\hyper@linkstart##1##2{}%
    \let\hyper@linkend\@empty\citet[#1][#2]{#3}}}      
 \newcommandtwoopt{\citeyearads}[3][][]%
   {\href{http://adsabs.harvard.edu/abs/#3}%
   {\def\hyper@linkstart##1##2{}%
    \let\hyper@linkend\@empty\citeyear[#1][#2]{#3}}}   
\makeatother

\newcommand{\rem}[1]{ } 

\title[Dark matter halos in the 2cDM model]
  {Dark matter halos in the multicomponent model. II. Density profiles of galactic halos}
\author[]
  {Keita Todoroki,$^1$\thanks{Email: keita@ku.edu}
  Mikhail V. Medvedev$^{1,2}$
  \newauthor 
             \\
  $^1$Department of Physics and Astronomy, University of Kansas, Lawrence, KS 66045\\
  $^2$Laboratory for Nuclear Science, Massachusetts Institute of Technology, Cambridge, MA 02139
  }
 \date{\today}

\pagerange{\pageref{firstpage}--\pageref{lastpage}} \pubyear{2016}

\def\LaTeX{L\kern-.36em\raise.3ex\hbox{a}\kern-.15em
    T\kern-.1667em\lower.7ex\hbox{E}\kern-.125emX}

\begin{document}


\label{firstpage}

\maketitle

\begin{abstract}
The multicomponent dark matter model with self-scattering and inter-conversions of species into one another is an alternative dark matter paradigm that is capable of resolving the long-standing problems of $\Lambda$CDM cosmology at small scales. 
In this paper, we have studied in detail the properties of dark matter halos with $M \sim 4-5 \times10^{11} M_{\odot}$ obtained in $N$-body cosmological simulations with the simplest two-component (2cDM) model. A large set of velocity-dependent cross-section prescriptions for elastic scattering and mass conversions, $\sigma_s(v)\propto v^{a_s}$ and $\sigma_c(v)\propto v^{a_c}$, has been explored and the results were compared with observational data. The results demonstrate that self-interactions with the cross-section per particle mass evaluated at $v=100$~km~s$^{-1}$ being in the range of $0.01\lesssim \sigma_0/m\lesssim 1$~cm$^2$g$^{-1}$ robustly suppress central cusps, thus resolving the core-cusp problem. The core radii are controlled by the values of $\sigma_0/m$ and the DM cross-section's velocity-dependent power-law indices $(a_s,a_c)$, but are largely insensitive to the species' mass degeneracy. These values are in full agreement with those resolving the substructure and too-big-to-fail problems. We have also studied the evolution of halos in the 2cDM model with cosmic time. 
\end{abstract}

\begin{keywords}
cosmology: theory -- methods: numerical -- self-interacting dark matter -- galaxies: formation
\end{keywords}


\section{Introduction}

The collisionless cold dark matter model (CDM) has become the dominant cosmological paradigm. It correctly represents the large scale structure of the universe but seems to be at odds with observations on small -- galactic and sub-galactic -- scales. Specifically, the problems are the core-cusp (CC) problem, the substructure (SS) problem and the too-big-to-fail (TBTF) problem. The CC problem deals with the $r^{-1}$ ``cusps'' of dark matter (DM) halo inner density profiles, seen in CDM simulations, but often not detected in observations which, instead, indicate smoother cores. The SS and TBTF problems are related to the paucity of dwarf galaxies with masses below about $10^{10}M_\odot$, which is not expected in the standard CDM and is not seen in numerical simulations. Various ideas have been proposed to settle the problems, including baryonic feedback, modified star formation, warm dark matter, self-interacting dark matter and others, but none so far convincingly resolve them across the full range of halo mass scales. A more detailed discussion of the problems and models is presented in our first paper (Paper 1) (Todoroki $\&$ Medvedev, in prep).

It has recently been suggested that a model of DM with more than one type of DM species, which allows for inelastic interactions between them, can resolve the above cosmological problems simultaneously \citep{medvedev2010, medvedev2010b, medvedev2014theo, medvedev2014}. In this series of papers, we are presenting numerical studies of the robustness of this model and explore constraints on the velocity-dependent cross-sections of elastic and non-elastic interactions. In Paper 1, we presented the comprehensive study of how the distribution of halos, i.e., the maximum circular velocity function (which is a better proxy of the halo mass function), is dependent upon $\sigma(v)$. Our results demonstrate that the 2cDM model with the self-interaction cross-section per particle mass, evaluated at $v=100$~km~s$^{-1}$, being in the range of $0.01\lesssim \sigma_0/m\lesssim 1$~cm$^2$g$^{-1}$ and the mass degeneracy of about $\Delta m/m\sim 10^{-7}-10^{-8}$ agrees well with observations. This indicates that the model can robustly resolve the SS and TBTF problems simultaneously by reducing the substructure with typical velocities below about 100~km~s$^{-1}$. In this paper, Paper 2, we explore the properties of dark matter halos: thair density profiles and evolution and how they are affected by various $\sigma(v)$-choices. We also present SIDM and CDM simulations for comparison. 

This paper is organized as follows. In Section \ref{sec:methods}, we briefly overview our multicomponent model and its numerical implementation. The studies of how the model parameters affect halo density profiles is presented in Section \ref{sec:profile_fit}. It also describes the fitting method used on the density profile. In Section \ref{sec:scaling}, we explore the scaling relations of the halo properties in comparison with the CDM and SIDM models. In Section \ref{sec:halo_evo}, we study the halo and halo core properties as a function of cosmic time. We also present velocity profiles in Section \ref{sec:Vprofiles}, for completeness. Finally, we summarize our conclusions in Section \ref{sec:CN}.


\section{Model and methods} \label{sec:methods}

The full description of the model and numerical methods are presented in Paper 1. 
Here we briefly overview the most important points. 

The two-component dark matter (2cDM) model is the simplest model in the general class of multicomponent DM ($N$cDM) models. It postulates that DM is comprised of two species: `heavy' and `light' mass eigenstates with masses $m_h>m_l$. Pair-wise interactions of these species is possible, which result in elastic scattering (no composition change) and inelastic conversion (i.e., the composition in the pair changes) of the species. The latter, inelastic, interactions result in the change of the total kinetic energy of the particles constituting the interacting pair by $\pm\Delta mc^2$ or $\pm2\Delta mc^2$ (where $\Delta m=m_h-m_l$), so that the total energy and momentum are conserved \citep{medvedev2010b, medvedev2014theo}. In this respect, the 2cDM model is kinematically similar to a large class of DM model with self-interactions and inelastic reactions. What differs 2cDM from them is that the quantum flavor-mixing of mass-eigenstates prevents destruction of heavier species in the early universe \citep{medvedev2014theo} -- the inevitable drawback of most multicomponent models with inelastic self-interactions. 

The inelastic processes in the flavor-mixed system is a unique quantum effect caused by the conversion of mass eigenstates of mixed particles in `seemingly elastic' interactions. If the velocities of the particles after such an interaction exceed the escape velocity of a dark matter halo, these particles escape. This effect was also referred to as `quantum evaporation' or the `M\"uncchausen effect'. This process leads to the gradual reduction, over the Hubble time, of small-scale halos below a certain critical mass. Higher mass halos are globally unaffected. However, the inner profiles of the halos are softened by interactions, similarly to the self-interacting dark matter. 

There are two main parameters in the 2cDM model, the mass degeneracy $\Delta m/m$ and the interaction cross-section per particle mass $\sigma/m$. The former characterizes the scale where the mass function and the maximum circular velocity functions have a break. The latter characterizes the strength of the effect of interactions. By comparing with observations, it has been shown that $\Delta m/m\sim 10^{-7}-10^{-8}$. In this study, we fix $\Delta m/m$ such that $V_k=c\sqrt{2\Delta m/m}=100$~km~s$^{-1}$, unless stated otherwise. 
We parametrize the velocity-dependent cross-section as 
\begin{equation} \label{eq:veldep}
 \sigma_{i\to f}(v) =     \left\{ \begin{array}{ll}
        \sigma_{0} (v/v_{0})^{a_{s}} & \mbox{for scattering,} \\ 
        \sigma_{0} (p_f/p_i) (v/v_{0})^{a_{c}} & \mbox{for conversion,} 
                \end{array}\right.
\end{equation}
where $p_i$ and $p_f$ are the initial and final momenta of the projectile particle, $v_{0}=100$~km/s is just the conventional velocity normalization and $\sigma_{0}$ is a model parameters that we vary. The power-law indices for scattering and conversion, $a_s$ and $a_c$, are also independent model parameters. 

Exploring the $\sigma_{0}/m$-$(a_s,a_c)$ parameter space is one of of the main goals of this paper. Among all the possibilities, $(a_{s}, a_{c})=(0,0)$ and $(-2,-2)$ are the best physically motivated: $(0,0)$ corresponds to the velocity-independent (hard-ball) cross-section, and $(0,0)$ is interesting because it naturally corresponds to the maximum conversion probability vs. scattering probability. The values of $a_{s}$ and $a_{c}$ were chosen to be $0,-1$, and $-2$ for both scattering and conversion and $-4$ for scattering only. The set of cross-sections was chosen to be $\sigma_{0}/m = 0.01, 0.1, 1$ and 10 cm$^{2}$g$^{-1}$. 

In simulations, we used the TreePM/SPH code GADGET \citep{springel2005, springel2008}, which was modified to include pairwise interactions of 2cDM, see Paper 1 for more detail. The initial composition, i.e., the total number of each DM species at the starting redshift, is taken to be 50:50, which is natural for fully decohered flavor-mixed particles. We use the Monte-Carlo technique to model DM-DM interactions under the assumption of {\em rare binary collisions} that any interaction probabilities are much smaller than unity at each timestep. The probabilities of the interaction processes which can occur within the time interval of $\Delta t$ are 
\begin{equation}\label{eq:probability}
P_{ij \rightarrow i'j'} = (\rho_{j}/m_{j})\sigma_{ij\rightarrow i'j'}|{\rm \bf v_{\it j} - v_{\it i}}| \Delta t \ \Theta(E_{i'j'}),
\end{equation}
where $i$ and $j$ denote a `projectile' and a `target' particle respectively, $\rho_{j}/m_{j}$ is the number density of the target particle, $\sigma_{ij\rightarrow i'j'}$ is the velocity-dependent DM cross-section for the process $ij\rightarrow i'j'$, where the unprime and prime signs denote initial and final states respectively, ${\rm \bf v_{\it j} - v_{\it i}}$ is the initial relative velocity of the interacting pair, and $\Theta(E_{i'j'})$ is the Heaviside function which dismisses kinematically forbidden processes. 

A periodic cube of side $3h^{-1}$ Mpc with the total number of $256^{3}$ particles was used, where {\it h} is the normalized Hubble constant, $h=H_0/(100~\textrm{km~s}^{-1}\textrm{Mpc}^{-1})$, and the initial force resolution was set to 0.6~kpc. Note also that the smoothing length  is allowed to be reduced up to 10$\%$ of the initial gravitational softening length during simulations. We are primarily interested in Milky Way-type or Andromeda-type halos so a small box of size $L = 3 h^{-1}$~Mpc was chosen to allow for better resolution of the halos. Thus, the largest and second-largest halos simulated in the box are on the order of $10^{11}$ M$_{\odot}$, and they are roughly a few factor smaller than the Milky Way and the Andromeda galaxies. In total, we performed over 57 simulations of 2cDM as well as several SIDM and CDM, for comparison. We assume the set of cosmological parameters consistent with \cite{planck2015}: $\Omega_{m} = 0.31$,  $\Omega_{\Lambda} = 0.69$,  $\Omega_{b} = 0.048$, $\sigma_{8} = 0.83$, $ n_{s} = 0.97$, $h = 0.67$.

The initial conditions were generated by the publicly available N-GenIC code with the same seed value to allow for reliable comparison across the models. The starting redshift was chosen to be $z_{i}=99$, and all simulations were carried to $z=0$.
We used the Amiga Halo Finder (AHF) \citep{knollmann2009} in post-processing analysis, e.g., to extract the halo properties and profiles.


 \begin{figure*}
  \centering
  \includegraphics[scale = 0.55]{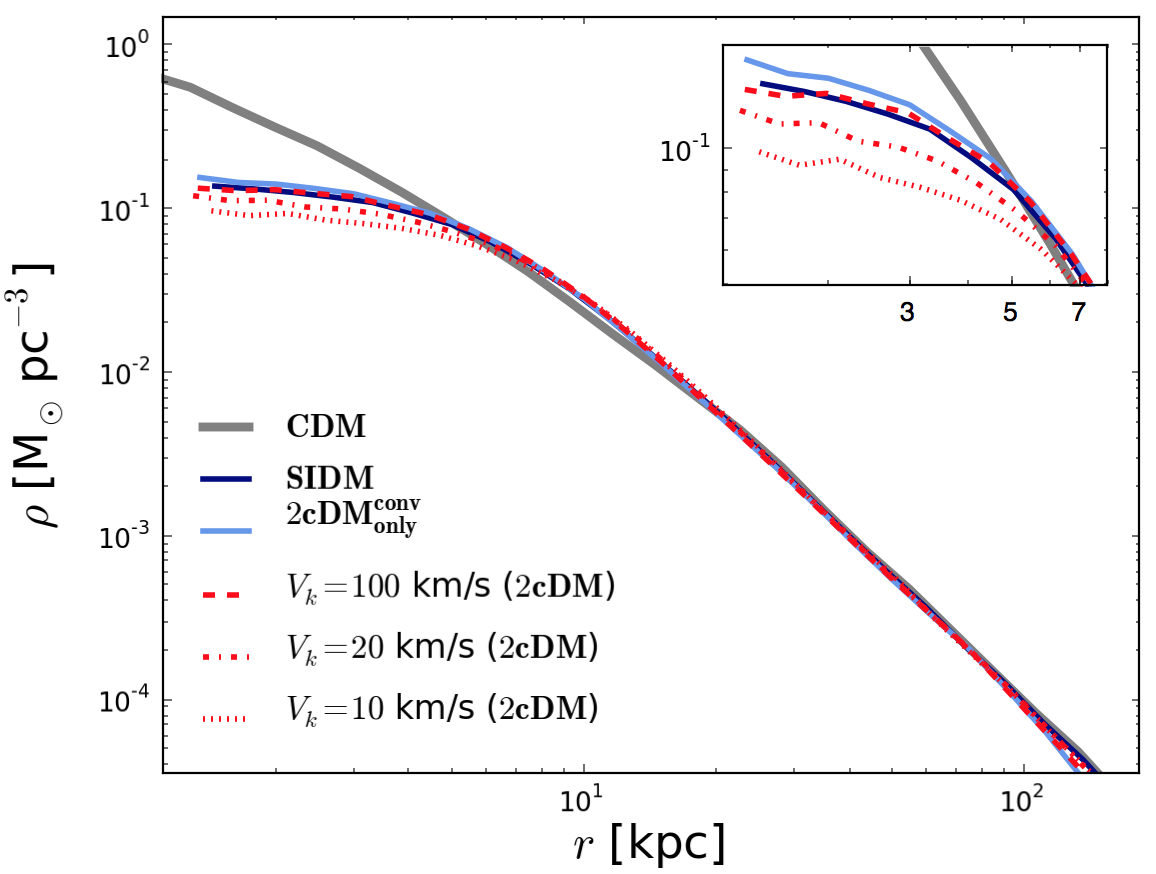}  
  \caption[]{\label{fig:profiles_SIDM} %
  DM density profiles for the classical CDM, SIDM and various 2cDM models with the fixed values of $\sigma_{0}/m = 1$ and $(a_{s}, a_{c}) = (-2,-2)$. 
  Here we compare the profiles of the most well resolved largest halo in the simulations that shows the general trend of the profile for each set of key parameter most clearly.
  The inset shows the inner-most region of the profiles to guide your eyes. 
  }
\end{figure*}

\begin{figure*}
  \centering
  \includegraphics[scale = 0.6]{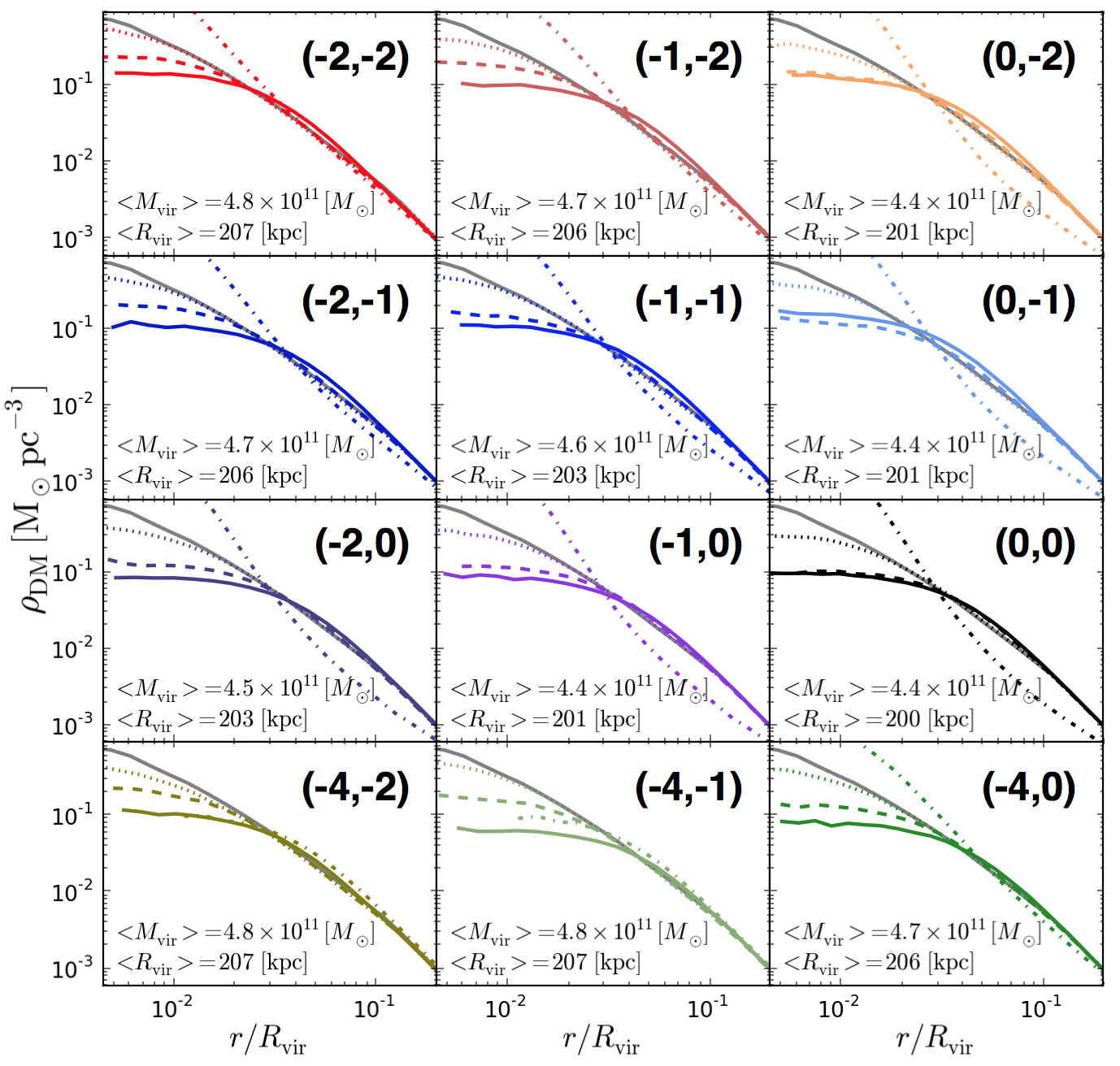}  
  \caption[]{\label{fig:profile_all} %
DM halo density profiles. Each label with a pair of numbers in parentheses represents a specific 2cDM model characterized by the elastic scattering and mass conversion power-law indices of the velocity-dependent DM cross-section, $(a_s, a_c)$ (Eq.~(\ref{eq:veldep})). The choice of the color scheme is arbitrary, but matches that of Figure~2 in Paper 1. Each panel shows the profiles of the largest halo in the simulated box of 3$h^{-1}$Mpc on a side with different DM cross-section values: $\sigma_{0}/m = 10$ (dotted-dash), 1 (solid), 0.1 (dash) and 0.01 (dotted) in the units of cm$^{2}$g$^{-1}$. The thick solid line is represents CDM simulations, for comparison.

  }
\end{figure*}

 \begin{figure}
  \centering
  \includegraphics[scale = 0.65]{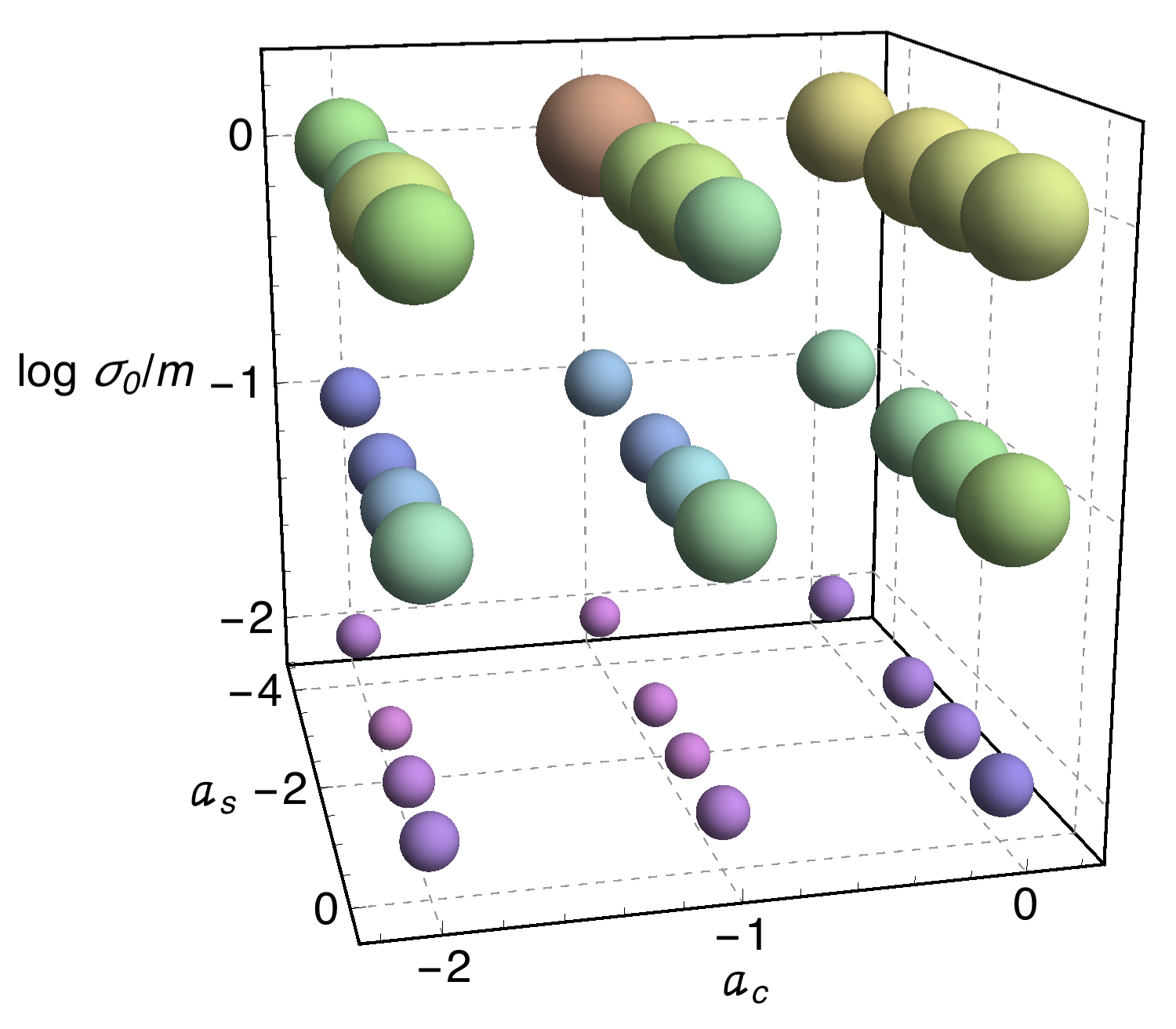}  
  \caption[]{\label{fig:sph} %
Visualization of the relative size of the core radii with respect to the DM cross-section, and the cross-section's velocity dependent exponents for scattering ($a_{s}$) and conversion ($a_{c}$). The color scheme used here is arbitrary, following gradations with red being the largest and violet being the smallest.
  }
\end{figure}

\begin{table}
\centering
\tabcolsep=0.11cm
\begin{tabular}{ccccccc}
\hline\hline
 Model & $\sigma_{0}/m$ [cm$^{2}$g$^{-1}$] & $r_{\rm c}$[kpc] & $\rho_{\rm c}$ [M$_{\odot}$pc$^{-3}$] &  $\beta$ \\
\hline

$(-2,-2)$ & 0.01 & 3.0 & 0.50 & 2.4 \\
 & 0.1 & 4.5 & 0.24 & 2.5  \\
 & 1 & 6.4 & 0.15 & 2.7 \\
 \hline
$(-1,-2)$ & 0.01 & 3.4 & 0.34 & 2.4\\
 & 0.1 & 5.0 & 0.18 & 2.6  \\
 & 1 & 7.6 & 0.10 & 2.8 \\
 \hline
 $(0,-2)$ & 0.01 & 3.6 & 0.30 & 2.5\\
 & 0.1 & 6.0  & 0.13 & 2.6 \\
 & 1 & 6.8 & 0.13 & 2.8 \\
 \hline
 $(-2,-1)$ & 0.01 & 3.1 & 0.40 & 2.4 \\
 & 0.1 & 4.8 & 0.19 & 2.5 \\
 & 1 & 7.3 & 0.11 & 2.8 \\
 \hline
 $(-1,-1)$ & 0.01 & 3.0 & 0.4 & 2.4\\
 & 0.1 & 5.4 & 0.15 & 2.6  \\
 & 1 & 7.3 & 0.11 & 2.8\\
 \hline
 $(0,-1)$ & 0.01 & 3.4 & 0.35 & 2.5\\
 & 0.1 & 6.3 & 0.12 & 2.6 \\
 & 1 & 6.2 & 0.15 & 2.8 \\
 \hline
 $(-2,0)$ & 0.01 & 3.6 & 0.31 & 2.5 \\
 & 0.1 & 6.2 & 0.12 & 2.6 \\
 & 1 & 8.0 & 0.081 & 2.7\\
 \hline
 $(-1,0)$ & 0.01 & 3.8 & 0.30 & 2.5\\
 & 0.1 & 6.5 & 0.12 & 2.6 \\
 & 1 & 7.8 & 0.083 & 2.7\\
 \hline
 $(0,0)$ & 0.01 & 4.1 & 0.27 & 2.5\\
 & 0.1 & 7.1 & 0.11 & 2.7 \\
 & 1 & 7.6 & 0.090 & 2.8\\
  \hline
 $(-4,-2)$ & 0.01 & 3.3 & 0.32 & 2.4\\
 & 0.1 & 4.4 & 0.20 & 2.5 \\
 & 1 & 6.6 & 0.11 & 2.7\\
  \hline
 $(-4,-1)$ & 0.01 & 3.1 & 0.40 & 2.4\\
 & 0.1 & 5.0 & 0.16 & 2.5 \\
 & 1 & 8.8 & 0.061 & 2.7\\
  \hline
 $(-4,0)$ & 0.01 & 3.6 & 0.33 & 2.5\\
 & 0.1 & 6.0 & 0.13 & 2.6 \\
 & 1 & 8.1 & 0.077 & 2.7\\[1ex]
\hline\hline 

\end{tabular}
\caption[]{ The gISO fit parameters for the 2cDM models for the density profiles of the largest halo. }
\label{table:fitting}
\end{table}

\section{Density profiles and fitting} \label{sec:profile_fit}

The most direct way to study the inner structure of the halo to address the CC problem is examine the density profiles. It is a known problem for nearly a few decades that the traditional CDM paradigm creates halos with `cuspy' or steep inner mass density profiles, which appear to be universal among a range of halo mass \cite[e.g][]{navarro1996, diemand2008}. Observations, on the other hand, have shown `cored' or nearly constant-density inner mass density profiles in many of the observed systems \citep{deblok2005, oh2008, kuzio2010, newman2009, newman2011}.

One of the favored solutions to the CC problem is the inclusion of baryonic physics with stellar feedback in the CDM paradigm instead of relying on exotic dark matter models. Successful creation of cored inner profiles has been reported by authors who found the core size of on the order of 1 kpc is strongly correlated to the stellar mass distribution where efficient transfer of stellar feedback energy occurs to dark matter \citep{pontzen2012, governato2012}. We note that this is a model-dependent result which requires a bursty, repeated star formation history and outflows. This, however, poses a question: How can a system that is deficient in stellar mass produce enough energy to affect and alter the more abundant (especially in dwarfs and low-surface brightness galaxies with $V_{\rm max} < 20$ km s$^{-1}$) dark matter distribution? Moreover, the cyclic and bursty baryonic feedback that can cause the formation of a small core of size $\sim 1$ kpc would clearly not be sufficient to resolve the TBTF problem, although the missing satellites or SS problem may be alleviated. \cite{papastergis2016} argued that the process of core creation itself does not resolve TBTF problem with the same reasoning. 

In contrast to these previous works, we show that DM-only simulations with our 2cDM model successfully indicate the presence of cored inner density profiles without relying on the baryonic physics. 
Figure~\ref{fig:profiles_SIDM} shows the density profiles of the largest halo in comparison with CDM, SIDM and 2cDM models, focusing specifically on studying how the elastic scattering alone (SIDM) and inelastic conversion only (2cDM$^{\rm conv}_{\rm only}$) play a role in forming a flat inner profile. 
We also compare a set of another key parameter in the 2cDM model, namely the kick velocity $V_{k}$ of 10, 20, and 100 km s$^{-1}$, to see its impact on the profiles.  
To ease the comparison, we use $(a_{s},a_{c}) = (-2,-2)$ as a fiducial model for SIDM, 2cDM$^{\rm conv}_{\rm only}$, and the full 2cDM cases. 
Firstly, by comparing SIDM and 2cDM$^{\rm conv}_{\rm only}$, we see that both scattering and conversion are independently and equally effective in reducing the central DM density and transforming a `cuspy' inner profile to a cored one. 
Secondly, although the difference among the profiles of the chosen values of $V_{k}$ appears to be small, it does show that smaller $V_{k}$ of 10 and 20 enhance the reduction of the central DM density compared to $V_{k} = 100$. The simple explanation is that $V_{k} = 10, 20$ is much smaller than the escape velocity of the halo, and so the interacted particles are subject to remain in the halo center. As a result, those particles trapped in the core would keep redistributing the DM particles' velocities, expanding the core size, reducing the central density, and thus further preventing the formation of a cuspy profile.

Figure~\ref{fig:profile_all} shows the density profiles of a variety of 2cDM cases, focusing now on exploring the physically and/or observationally motivated parameter set of $(a_{s},a_{c})$ and $\sigma_{0}/m$, and comparing with that of the CDM counterpart. We find that the degree of flatness of the inner profile strongly depends on the magnitude of DM cross-section, but the differences seen among the different sets of velocity-dependent power-law indices ($a_{s}, a_{c}$) are relatively small. The general trend is as follows. The larger the cross-section, the larger the core and the smaller the central density become. 
We quantified this relation through fitting, which is described in details in the next section. This trend is expected from the fact that a larger cross-section produces more DM-DM interactions, which deliver more thermal energy from the outer to the central region of the halo. Another general trend that is common to nearly all the cases is that the effect of mass conversions seems to be less pronounced as compared to that of elastic scattering. To see it clearly, it is helpful to compare the panels column by column and row by row. 
Then, one sees minor differences in profiles obtained with different $a_{c}$ while holding $a_{s}$ constant, i.e., within one column, but slightly more differences between the columns.
Interestingly, we see the cases of $\sigma_{0}/m = 1$ and 0.1 with $a_{s} = 0$ to produce very similar profiles. 

More notably, the largest cross-section of $\sigma_{0}/m = 10$ consistently shows steeper inner profiles for nearly all the cases -- even steeper than the CDM counterpart. This is attributed to gravothermal collapse triggered by the excessive amount of DM-DM interactions in the inner region of the halo. 

Gravitating systems possess negative heat capacity, and that heating, i.e., increasing its average kinetic energy (or temperature), $\langle K\rangle$, reduces its total energy, $E=K+U$, where $U$ is the potential energy. This is easily seen, for example, from the virial equilibrium relation for a central potential with $U\propto r^{-\gamma}$ that $2\langle K\rangle =-\gamma \langle U\rangle$, so that $E=-(2/\gamma-1)\langle K\rangle<0$ for gravitating systems with $0<\gamma<2$.
Thus, if a halo core contracts slightly, it adiabatically heats up and, thus, loses energy. This triggers collisional heat transport (if collisions are present), which further heats up the interior leading to even stronger energy loss. This runaway process, called ``gravothermal catastrophe'' or ``gravothermal collapse'', occurs on the time-scale of the order of tens of dynamical times of the halo core \citep{lynden-bell1980,binney2008}. Ultimately, the system reaches thermal equilibrium due to collisions and forms an isothermal sphere with the cuspy $\rho(r)\propto r^{-2}$ profile. The presence of mass conversions, which can remove particles from a halo, can affect the evolution as well as the final state. Note that fully collisionless systems, e.g., CDM halos, are stable. 

In 2cDM (and SIDM) simulations, a collisionless cuspy Navarro-Frank-White (NFW) profile \citep{navarro1996,navarro1997} forms first\footnote{It turned out that this is only true when the DM cross-section's velocity dependence is weaker with smaller $\sigma_{0}/m$ values. For a stronger velocity dependence, such as ($-2,-2$), halos tend to form non-cuspy, flat inner profile at least as early as $z > 10$.}. Then, on the time scale of elastic DM-DM collisions, heat from the outer regions is transported inwards, thus heating up the interior and establishing an isothermal core. On an even longer time-scale, the collapse sets in to produce the isothermal sphere.

The isothermal core collapse is precisely what is seen in 2cDM (and SIDM) simulations with $\sigma_{0}/m = 10$. Although, all collisional systems, $\sigma_{0}/m \not= 0$, are subject to the collapse, the rate may exceed the Hubble time, provided $\sigma_{0}/m$ is small enough. That is why halos in simulations with $\sigma_{0}/m$ of order unity or less are not collapsed. 

Lastly, we should note that the three cases with $a_{s} = -4$ may not be fully trustworthy for numerical reasons. We find that by $z=0$ a substantial fraction of all the DM particles experience their interaction probabilities $P_{ij\to i'j'}$ (see Eq. (\ref{eq:probability})) in excess of 0.1. Consequently, the ones in the cores of the gravitational potentials are no longer in the weak collision regime, where Eq. (\ref{eq:probability}) is valid. We note that this is true for all the  cases with $a_{s} = -4$ and $\sigma_{0}/m = 10$. Accurate study of these cases requires better resolution simulations.

\subsection{Fitting for the density profile}

The conventional CDM halos are known to possess the NFW profile, and its generalized form, or the generalized NFW profile (gNFW, \citet{zhao1996}) is written as
\begin{equation}
\rho_{\rm gNFW}(r) = \rho_{s}  \left( \frac{ r }{ r_{s}} \right)^{-\alpha} \left( 1+ \frac{ r }{r_{s}} \right)^{\alpha - \beta}.
\end{equation}
With the broken power-law profile with $\alpha$ and $\beta$ being the inner and outer power-law indices, the NFW profile, which takes $\alpha=1$ and $\beta=3$.

One of the key outcomes of the 2cDM physics is the creation of a cored inner profile, deviating from the cuspy CDM distribution well presented by the universal NFW profile, or perhaps the Einasto profile, which can be viewed as a better alternative \citep{springel2008, merritt2006,ludlow2016,klypin2016}. 
Figure~\ref{fig:profile_all} illustrates clearly that with 2cDM, even the cross-section as small as $\sigma_{0}/m = 0.01$ is capable of producing a small-sized cored inner profile. 
For this reason, we sought for an alternative fitting model to the NFW profile. The Einasto profile is also not so appealing in the current study since NFW and Einasto profiles are only mildly different within the range of $r$ considered. The larger number of parameters used in the Einasto profile (total of five) is also unappealing as it reduces statistical significance of the fits. 

A more observationally-motivated alternative is the so-called spherical pseudo-isothermal halo profile (ISO) that has been a favored fitting function used by some observers for LSB galaxies \citep{deblok2002,swaters2003,deblok2005,kuzio2010}:

\begin{equation}
\rho_{\rm ISO}(r) = \rho_{\rm c}  \left[1+  \left(\frac{r}{r_{\rm c}} \right)^{2} \right]^{-1},
\end{equation}
where $r_{\rm c}$ and $\rho_{\rm c}$ are the radius and the density of the core respectively. 
Similarly, the Burkert profile, a phenomenologically derived profile \citep{burkert1995, dalcanton2001, salucci2007, salucci2012} can describe the constant inner density, $\rho \sim \rho_{c}$, at $r \ll r_{c}$ just like ISO, but only differs at $r \gg r_{c}$ where $\rho \propto r^{-2}$ for ISO and $\rho \propto r^{-3}$ for Burkert. 
These models make a clear distinction between the inner cored and the outer steeply declining profiles.
Note that in these models, the outer power-law index is not parametrized but set as a constant value, thus lacking the flexibility in adjusting the outer slope.

The assumption of the spherical isothermal central halo manifestly represented by the ISO is fairly reasonable and robust, given the fact that nearly all the cases of the 2cDM simulations exhibit distinctive cored inner profiles. Thus, the ISO (or Burkert) model can accurately captures the core size and density as long as the inner radial region is well-resolved. However, the ISO yields a very poor fit to the outer halo, owning to the fixed outer slope of $-2$, which is not generally seen. 
The Burkert model fits better to the 2cDM profiles than the ISO, yet its accuracy is compromised for cases with 'mildly' cored profiles, or cases with smaller core sizes where our simulations do not have enough resolution to probe for. 
To mitigate this problem, we propose a generalized ISO (gISO) profile as follows:
\begin{equation}
\rho_{\rm gISO}(r) = \rho_{\rm c} \left[1+ \left(\frac{r}{r_{\rm c}}\right)^{2} \right]^{-\beta/2}.
\label{gISO}
\end{equation}
By introducing the parameter $\beta$ in order to account for the steeper slope of the outer density profile, we find that the gISO model fits remarkably well with the cored density profiles over the entire radial range our simulations were able to resolve, see figures in Section \ref{sec:halo_evo}. Introducing the $\beta$ parameter also proved useful to quantify the discrepancy between the pure ISO and gISO; note that gISO with $\beta=2$ becomes the standard ISO.

Here we present the results of fits of the largest --- and hence best resolved --- halo ($N_{\rm tot} > 2$ mill. within the virial radius) in each simulation with various values of parameters $(a_{s}, a_{c})$ and $\sigma_0/m$.
We remind the reader that for all those cases in comparison we used the exactly the same initial condition, thereby we conducted the fit on the same halo but simply with different parameter settings.
Table~\ref{table:fitting} lists the core radii, core densities and $\beta$ obtained from the gISO fit for `good' cored cases, which exclude $\sigma_{0}/m = 10$ cases for reasons explained earlier. Overall, our fitting results clearly show a good consistency in capturing the general trend: We find that a larger DM cross-section corresponds to  (i) a larger core radius, (ii) a smaller core density, and (iii) a larger $\beta$, regardless of the set of parameters ($a_{s}, a_{c}$), i.e., the DM cross-section's velocity dependence. Note that the values of $\rho_c$ and $r_c$ reported in Table~\ref{table:fitting} are in great agreement with those derived from observations by  \cite{deblok2005} who found the range of halo core densities and radii from LSB galaxies (with the ISO fit) to be 0.01--1 M$_{\odot}$pc$^{-3}$ and 0.27--6.8 kpc, respectively.

Figure~\ref{fig:sph} visualizes the relative size of the core radii with respect to the DM cross-section and the set of parameters for $(a_{s}, a_{c})$. It indicates that, aside from the general trend described above, $a_{s} = 0$ produces slightly larger core radii than that of $a_{s} = -1,-2$ or $-4$. Overall, the core sizes are generally smaller for more negative values of $a_{s}$ and $a_{c}$. This trend is better seen for smaller cross-sections $\sigma_{0}/m = 0.1$ and 0.01. We caution again that $a_{s} = -4$ may not be as trustworthy because of the poorer fit given by gISO (especially for $\sigma_{0}/m = 0.01$), although not significantly so, possibly due to the numerical effects discussed above.

\begin{figure}
  \centering
  \includegraphics[scale = 0.42]{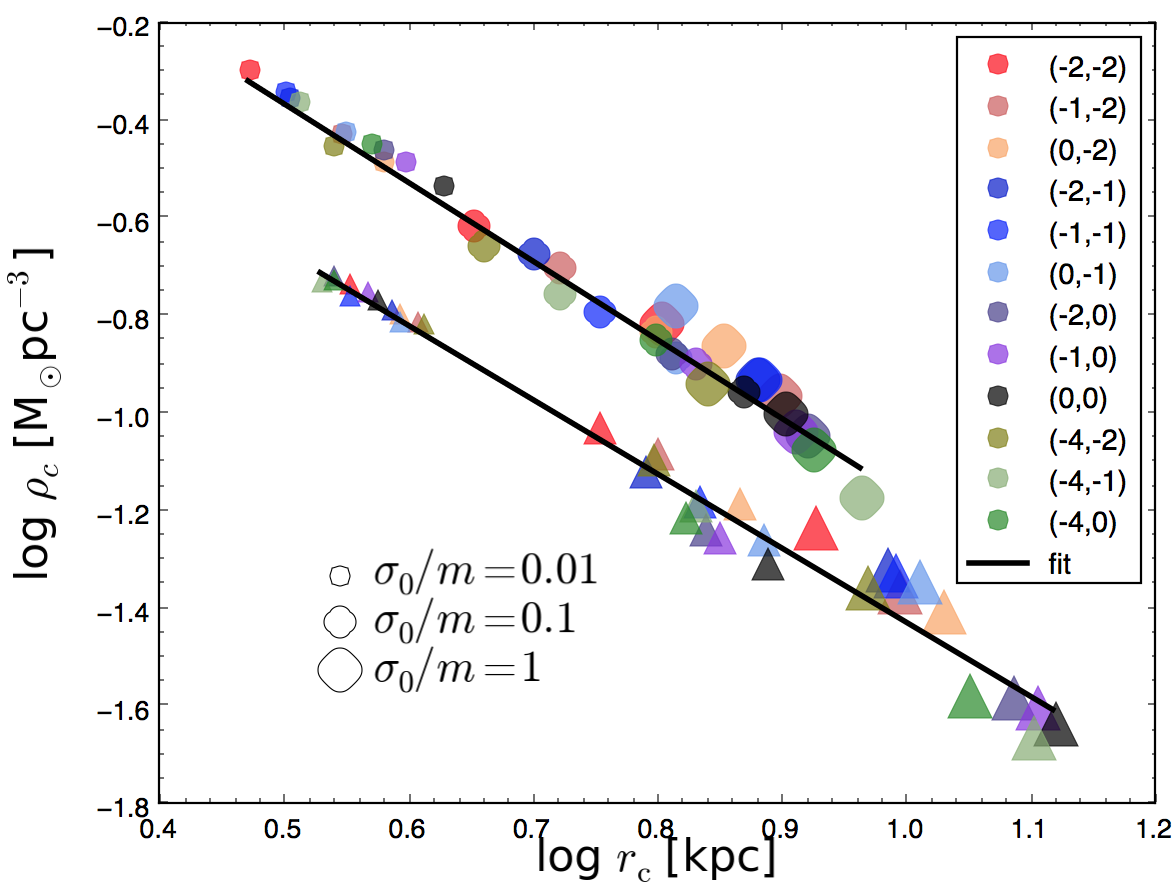}  
  \caption[]{\label{fig:scaling-rc-rhoc} %
Core densities versus core radii for different models. The largest (circles) and second largest (triangles) halos with their virial masses $M_{\rm vir} \sim 5 \times 10^{11}$ and $4 \times 10^{11}$M$_{\odot}$ in the simulation are plotted. 
  }
\end{figure}

 
\section{Halo scaling relations} \label{sec:scaling}

In this section we explore the scaling relations obtained from the gISO fits and study the time evolution of the halo properties. We first show how the fit parameters are correlated with each other for different sets of ($a_{s}, a_{c}$). Figure~\ref{fig:scaling-rc-rhoc} is a scatter plot of the core density vs. core radius of the largest and the second largest halos with their virial masses $M_{\rm vir} \sim 5 \times 10^{11}$ and $4 \times 10^{11}$M$_{\odot}$ at $z = 0$ in the simulation. We stress that these two halos can be considered 'quasi-quiescent', meaning that both follow an identical halo evolution history with no signs of major merger events in the late time. 
From the figure, one can see a very tight correlation, which can be fit with power laws:
\begin{equation}
\begin{split}
\log  \ \rho_{c} &= 0.44 -1.6 \ \log  \ r_{c} \ ({\rm largest \ halo}), \\
\log  \ \rho_{c} &= 0.095 -1.5 \ \log  \ r_{c} \ ({\rm second \ largest \ halo}).
\end{split}
\end{equation}
From the above equations, we can see that the slopes for the two largest halos are very similar, indicating a nearly universal dependence 
\begin{equation}
\rho_c\propto r_c^{-1.5}.
\end{equation}
Note, this scaling only holds for the particularly selected halos of $\sim 10^{11}$M$_{\odot}$ with varying $\sigma_0/m$. The fit among different halos with different mass scales with a fixed value of $\sigma_0/m$ would result in a different scaling.
The offset of the trend lines in the figure is clearly due to the difference in the mass of the halos. 

It is remarkable that the above generic negative power-law dependence we have found, has also been deduced from observational data \citep{burkert1995, kravtsov1998, barnes2004, plana2010, salucci2007, spano2008, salucci2012, kormendy2004, kormendy2016}. The types of the systems studied in these papers range from dSphs to galaxy clusters. Thus, the trend that the smaller core radius, the larger central density seems to be indeed universal. Our results with the slope of $\sim -1.5$ appear to be in approximate agreement with the results by \cite{kormendy2004}, for example, who used the pure ISO fits on their samples consisting of the late-type to dSph galaxies yielding the slope of $\sim -1.0$ to $-1.2$. Similarly, \cite{plana2010} applied the pure ISO fit to a sample of observed late-type galaxies and obtained the slope to be ranging $\sim -1.0$ to $-1.6$. Thus, these observational results are consistent with our results, especially taking into account that the ISO fit model used elsewhere appears to not be a perfect fit to the simulated profiles.

\begin{figure*}
  \centering
  \includegraphics[scale = 0.58]{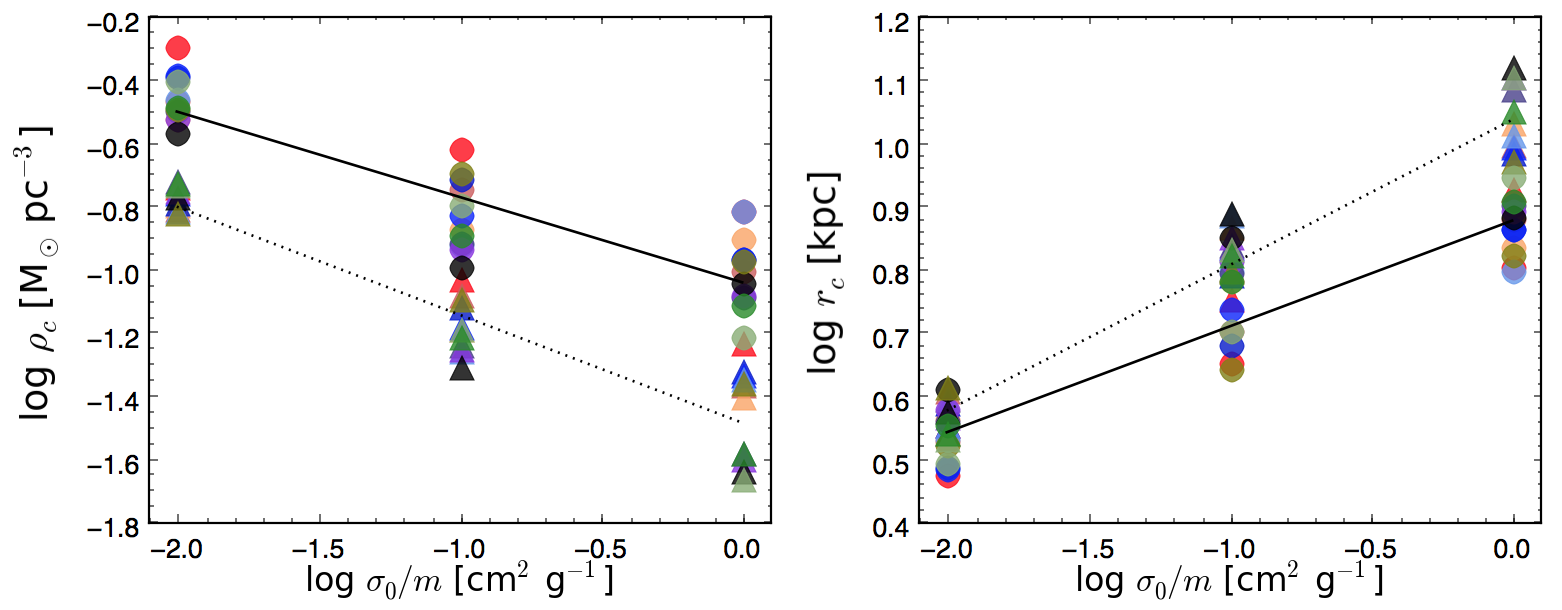}  
  \caption[]{\label{fig:explore-sigma} %
Core densities and core radii versus the DM cross-section, $\sigma_0/m$. Same as Figure~\ref{fig:scaling-rc-rhoc}, the circles and triangles represent the largest and second-largest halos, and it follows the same color scheme. The solid line is the fit to the largest halo and the dashed line is for the second-largest halo. 
  }
\end{figure*}

We also explored the scaling relations between halo core parameters ($\rho_{c}$ and $r_{c}$) as a function of the DM cross-section $\sigma_{0}/m$. Figure~\ref{fig:explore-sigma} quantitatively confirms what is discussed earlier: larger cross-sections result in larger cores with smaller core densities. The scaling relations obtained from the gISO fits are 
\begin{equation}
\begin{split}
\log  \ \rho_{c} &= -1.0 -0.27 \ \log  \ (\sigma_{0}/m) \ ({\rm largest}), \\
\log  \ \rho_{c} &= -1.5 -0.34 \ \log  \ (\sigma_{0}/m) \ ({\rm second \ largest})
\end{split}
\end{equation}
and
\begin{equation} \label{eq:rc_sigma}
\begin{split} 
\log  \ r_{c} &= 0.88 + 0.17 \ \log  \ (\sigma_{0}/m)  \ ({\rm largest}), \\
\log  \ r_{c} &= 1.0 + 0.23 \ \log  \ (\sigma_{0}/m) \ ({\rm second \ largest}).
\end{split}
\end{equation}
The scatter is larger in these scalings among different set of ($a_{s}, a_{c}$), the fits show that the slopes for both $\rho_{c}$ and $r_{c}$ versus $\sigma_{0}/m$ are consistent with each other with the approximate dependencies being
\begin{equation}
\rho_c\propto (\sigma_0/m)^{-0.3}, 
\quad
r_c\propto (\sigma_0/m)^{0.2}.
\end{equation}

\begin{figure}
  \centering
  \includegraphics[scale = 0.53]{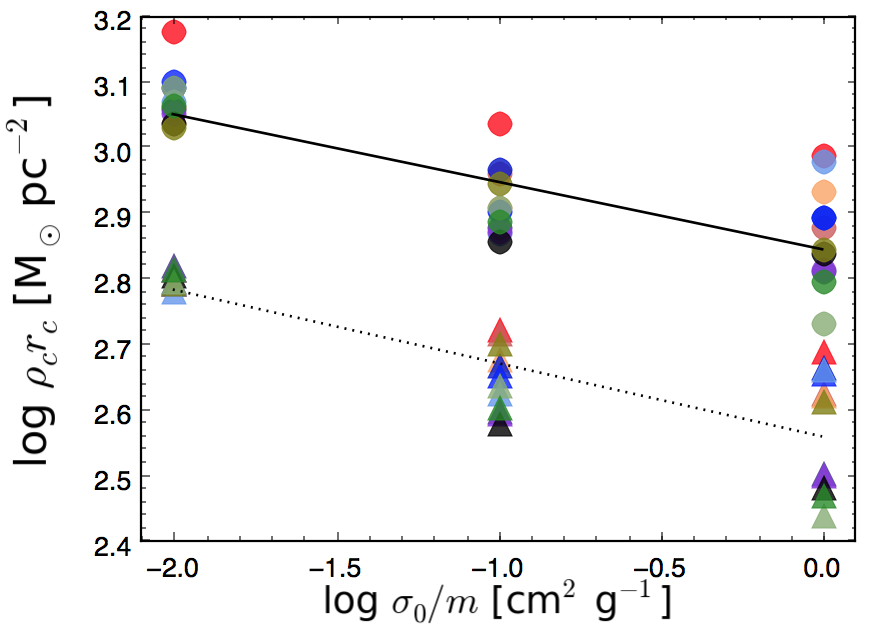}  
  \caption[]{\label{fig:rcrhoc_sigma} %
The surface DM density, or the product of core densities and core radii, versus the DM cross-section.
  }
\end{figure}

\begin{figure}
  \centering
  \includegraphics[scale = 0.48]{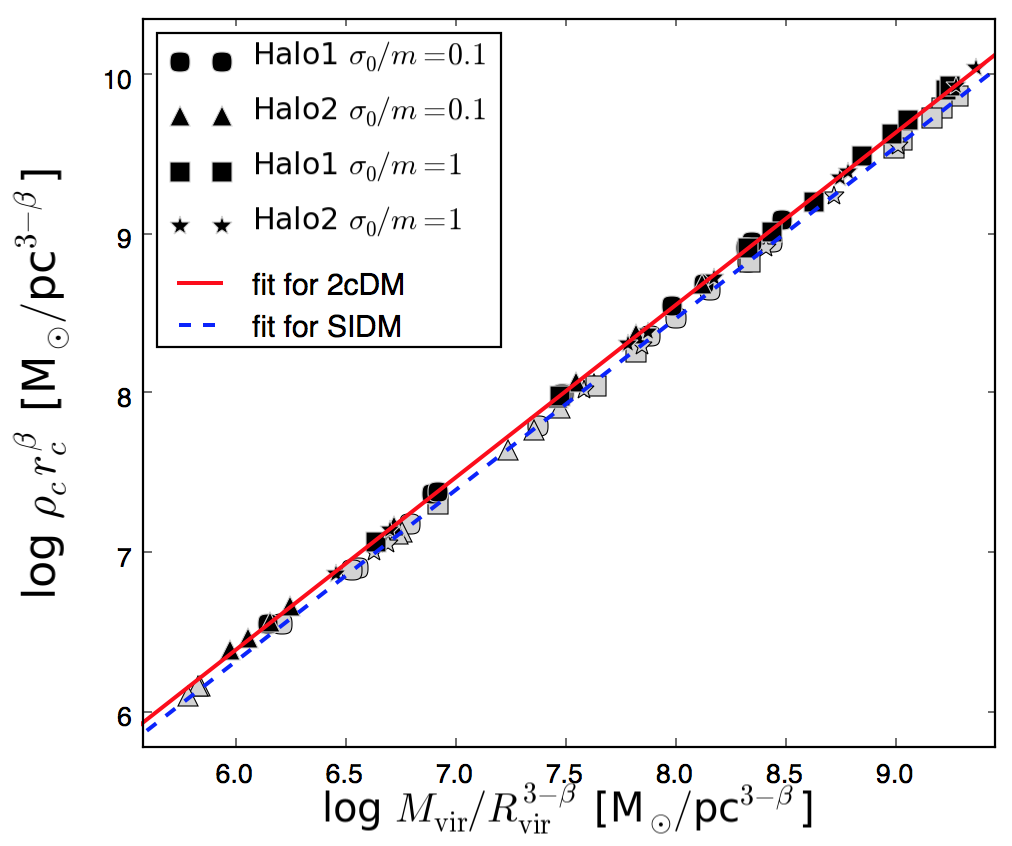}  
  \caption[]{\label{fig:page7} %
Time dependent scaling relation between the halo core parameters and the halo virial properties. 
The largest and second largest halos ('Halo1' and 'Halo2') for $\sigma_{0}/m = 0.1$ and 1 are shown for 2cDM (black) and SIDM (gray). 
Total 9 points are taken for each case at different redshifts from $ z \sim 3$ to 0 and  are presented without distinguishing them by color or symbol shape. The solid and dashed lines are the fits for 2cDM and SIDM, respectively.

  }
\end{figure}

Interestingly, the observationally derived product of $\rho_{c}r_{c}$ (i.e., the surface DM density), has been shown to be approximately constant and ubiquitous among different types of galaxies over a wide range of absolute magnitudes \citep{spano2008, gentile2009}. In Figure~\ref{fig:rcrhoc_sigma} we show the scaling relation of $\rho_{c}r_{c}$ versus $\sigma_{0}/m$ for the different set of ($a_{s}, a_{c}$). Here we only show the largest (circles) and second-largest (triangles) halos in the simulation with a minor difference in their halo masses. However, the slopes are remarkably shallow, 
$\rho_{c} r_{c} \propto (\sigma_{0}/m)^{-0.1}$,
with a small offset in the magnitude of $\rho_{c} r_{c}$. This indicates consistency with observational results, especially given large scatter across the simulated sample:
\begin{equation}
\begin{split}
\log  \ \rho_{c} r_{c} &= 2.8 - 0.1 \ \log  \ (\sigma_{0}/m) \ ({\rm largest}), \\
\log  \ \rho_{c} r_{c} &= 2.6 - 0.1 \ \log  \ (\sigma_{0}/m) \ ({\rm second \ largest}).
\end{split}
\end{equation}
The statistical significance of the small deviation from $\rho_{c}r_{c}\sim const.$ seen in the figure can be examined by requiring a larger sample of halos without compromising the numerical resolution effects. In addition, we also would need to account for the baryonic processes, which play a significant role in forming the observed correlation of $\rho_{c}$ and $r_{c}$. These issues are planned to be explored in forthcoming studies.

Given that the DM density profiles are well described by gISO given by Eq. (\ref{gISO}), a useful relation between the core and virial parameters can be derived. The virial mass is the sum of the masses, $M_{c}$ and $M_{out}$, of the constant density core and the outer region where $\rho\propto r^{-\beta}$. Thus, 

\begin{equation}
\begin{split}
M_{\rm vir} & = M_{c} + M_{out} \\
 & \simeq \int^{r_{c}}_{0} \rho_{c} 4 \pi r^{2} dr  + \int^{R_{\rm vir}}_{r_{c}} \rho_{c} \left(\frac{r}{r_{c}} \right)^{-\beta} 4 \pi r^{2} dr,
\end{split}
\end{equation}
where $0<\beta<3$, as follows from simulations. This yields
\begin{equation}
M_{\rm vir} = 4 \pi r_{c}^3 \rho_{c} \left[ \frac{1}{3-\beta} \left(\frac{R_{\rm vir}}{r_{c}} \right)^{3-\beta} - \frac{\beta}{3(3-\beta)} \right]
\end{equation}
Since $R_{\rm vir} \gg r_{c}$, the second term in the above equation can be neglected. This relation should hold at all times, even if the masses, radii and $\beta$ are functions of time, $t$. Thus, we have a time-dependent scaling relation 
\begin{equation}\label{eq:page7}
\rho_{c}(t) r_{c}(t)^{\beta(t)} \propto \frac{M_{\rm vir}(t)}{R_{\rm vir}^{3 - \beta (t)}(t)}.
\end{equation}

Figure~\ref{fig:page7} shows that Eq.(\ref{eq:page7}) is indeed very robust. Here we present both 2cDM and SIDM results in order to ease the comparison of the two models, which differ in that SIDM has elastic scatterings between DM particles but no mass conversions, whereas 2cDM has both. In other words, any difference that would be seen between the 2cDM and SIDM results is due to the inelastic mass conversions. The figures illustrate the results for the largest and second-largest halos with $\sigma_{0}/m = 0.1$ and $1$ as example cases. Obviously, the accuracy of Eq.(\ref{eq:page7}) manifests the fact that both SIDM and 2cDM halos are well represented by the gISO profile regardless of the specific choice of the parameters ($a_{s}, a_{c}$) and $\sigma_{0}/m$ used in this work. The log-log fits show a remarkable similarity between 2cDM and SIDM as well:

\begin{equation}
\begin{split}
\log \rho_{c}(t) r_{c}(t)^{\beta(t)} = - 0.10 + 1.1 \ \log \frac{M_{\rm vir}(t)}{R_{\rm vir}^{3 - \beta (t)}(t) }, \ \ \text{(2cDM)} \\
\log \rho_{c}(t) r_{c}(t)^{\beta(t)} = - 0.14 + 1.1 \ \log \frac{M_{\rm vir}(t)}{R_{\rm vir}^{3 - \beta (t)}(t)}. \ \ \text{(SIDM)}
\end{split}
\end{equation}


\begin{figure*}
  \centering
  \includegraphics[scale = 0.55]{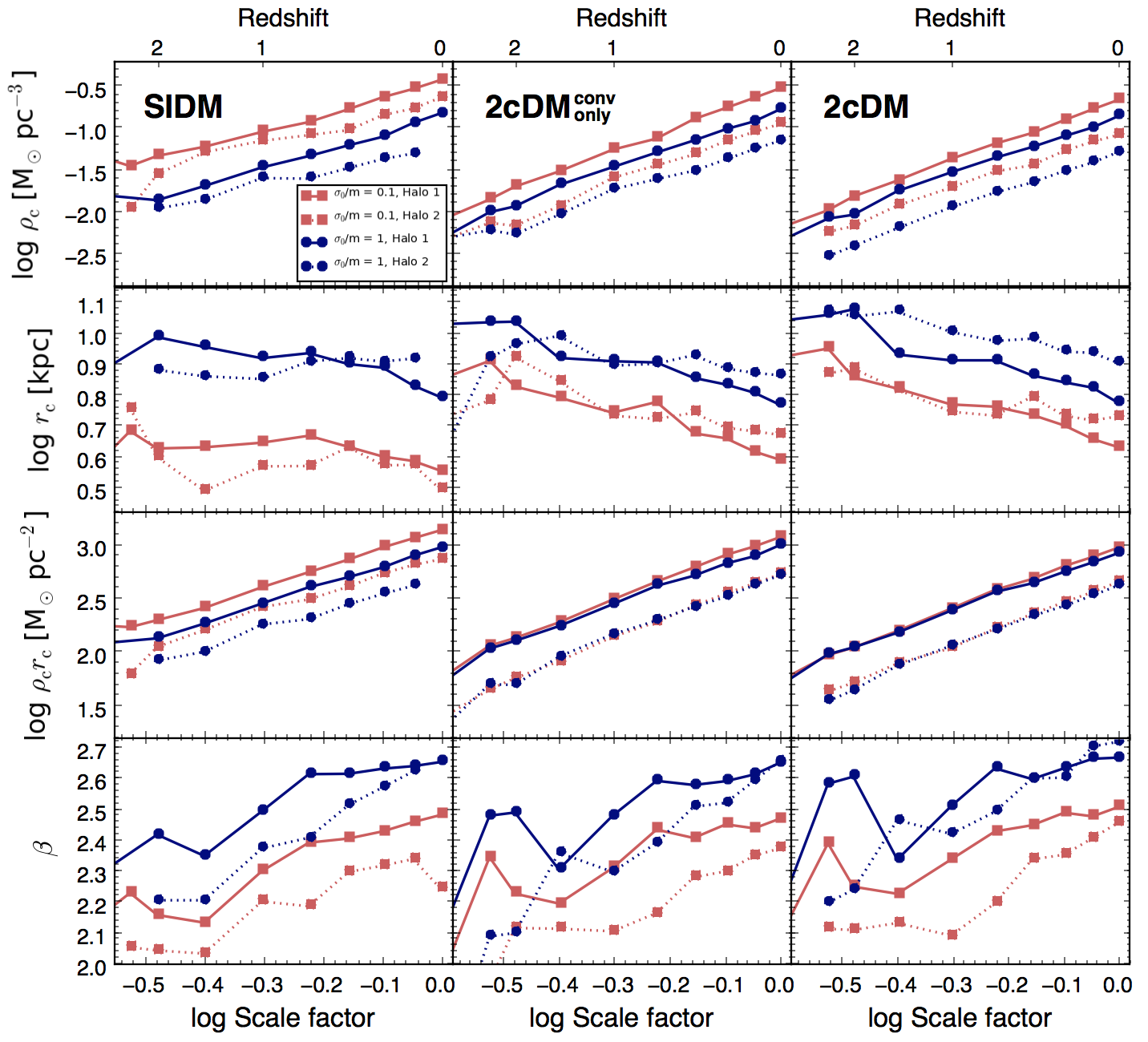}  
  \caption[]{\label{fig:halocore_evo} %
Halo core fitting parameters versus scale factor. 
The largest (solid) and second largest halos (dotted) ('Halo1' and 'Halo2') are shown and the results for $\sigma_{0}/m = 0.1$ (red) and 1 (blue) are compared.
All the cases, except for $\beta$, were fitted with a power-law and the results are summarized in Table~\ref{table:halocore_evo} in Appendix.
  }
\end{figure*}

\section{Halo evolution} \label{sec:halo_evo}

In this section, we study how the halo core and halo itself evolve over time. In order to isolate the effects of elastic scattering and inelastic mass conversions of DM particles, we compare three models (i) 2cDM, (ii) 2cDM$^{\rm conv}_{\rm only}$ (i.e., `conversion only', no scattering) and (iii) SIDM (i.e., `scattering only').
We present $\sigma_{0}/m = 1$ and 0.1 cases only, because the profiles obtained in our $\sigma_{0}/m = 0.01$ simulations tend to show only a mild reduction of the central density within the inner radial range our simulations were able to resolve, which degrades the goodness of the gISO fits.
As before, only the results for the largest and second-largest halos in the simulation box are presented. The third and fourth largest halos were also checked. However, they appear to experience some major mergers at later times, which introduce much non-linearity in the DM interaction processes, and hence in the halo evolution, because these rates scale with density and velocity as $\propto\rho^2\sigma(v) v$. To ease the comparison and for simplicity, we therefore do not present these cases here. Since it appears that all the fitting parameters generally follow a power-law, we summarized the fitting results quantitatively in Table~\ref{table:halocore_evo} and \ref{table:halo_evo} in Appendix. To minimize the numerical resolution errors, we chose the range of the universe scale factor of $0.25 \leq a \leq 1$ ($z=3$ to 0). We use the ($-2,-2$) case as the fiducial for SIDM, 2cDM$^{\rm conv}_{\rm only}$, and 2cDM hereafter.

\subsection{Halo core properties \emph{vs.} scale factor}

In Figure~\ref{fig:halocore_evo}, we show the time evolution of the fitting parameters, namely $\rho_{c}$, $r_{c}$, $\rho_{c}r_{c}$ and $\beta$, as a function of scale factor. One can see that the core density versus scale factor follows a tight power-law relation for both 2cDM and SIDM. The time evolution of the core radii, however, shows a less tighter power-law with a mildly declining slope for the cases presented in the figure. We attribute this `core compression' to the growth of halo mass due to DM accretion. 
One would expect that at the onset of a halo evolution a NFW-like halo forms on the dynamical time scale, if the velocity-dependence of the DM cross-section is relatively weak and the cross-section value is small. Later on, a halo core appears and then its size increases with time due to DM collisions, which happens on the much longer collisional time. Once the core has formed, the subsequent evolution is an adiabatic change of the collisionless equilibrium caused by gradual accumulation of mass, provided no mergers occur. 
This accumulation of mass is then followed by a continuous increase in the central density and a mild reduction in the core size toward the present time. 
We find such evolution is universal across 2cDM and SIDM models. 
In general, the difference in the power-law relations between $\sigma_{0}/m = 0.1$ and 1, and the two most well-resolved halos are minimal. 
Quantitatively, for both 2cDM$^{\rm conv}_{\rm only}$ and 2cDM we have roughly $\rho_{c} \propto a^{2.4}$, $r_{c} \propto a^{-0.4}$, and $\rho_{c} r_{c} \propto a^2$ over the period of $0.25 \leq a \leq 1$ where $a$ is the universe scale factor (Table~\ref{table:halocore_evo}). 

It is also instructive to look at the time evolution of the DM core surface density (i.e., the product of the core density and core radius $\rho_{c}r_{c}$), which shows an even tighter power-law relation among the cases with different $\sigma_{0}/m$, especially for 2cDM. By comparing the three models, we find that mass conversions are playing an important role in halo core evolution.  
Lastly, we show how the outer density power-law index $\beta$ evolves as a function of the scale factor in the bottom row panels. The most striking is that, in general, $\beta$ increases from $\sim$2.0, indicative of an isothermal halo, at $z=3$ to $\sim$2.7 at $z=0$. This is likely because of a fairly rapid mass accumulation, so that the rate of DM interactions is too slow and thus unable to equilibrate temperature throughout the halo. Whether the other models, such as (0,0) other than ($-2,-2$) yields the same evolutionary trend is not explored in this work.

\subsection{Global halo properties \emph{vs.} scale factor}

\begin{figure*}
  \centering
  \includegraphics[scale = 0.55]{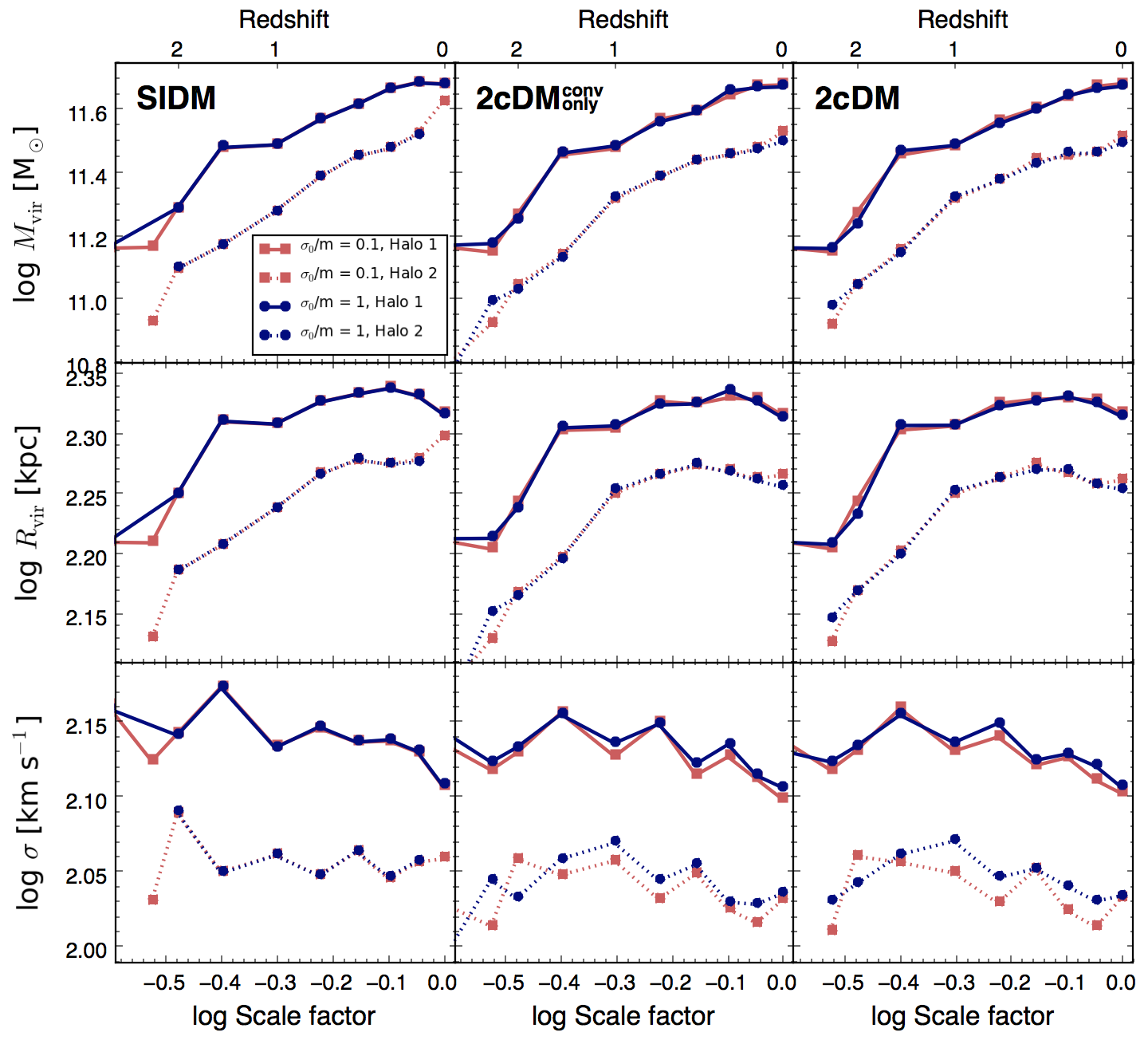}  
  \caption[]{\label{fig:halo_evo} %
Global halo properties (the virial mass, virial radius and the velocity dispersion) versus scale factor. The line styles and colors are the same as Figure~\ref{fig:halocore_evo}. The fit parameters to each data set are presented in Table~\ref{table:halo_evo} in Appendix.
  }
\end{figure*}

In Figure~\ref{fig:halo_evo} we show the time evolution of the global halo properties, such as the virial mass and radius, as well as the velocity dispersion. It is clear to see that the virial mass of the halo grows approximately as a power-law, in contrast to the virial radius of which growth appears to slow down at the late times (roughly $z \lesssim 1.5$). Apparently, this is consistent with the positive slope in the $\log \rho_{c}$--$\log a$ plot shown earlier (again, here $a$ without a subscript is the scale factor, not the exponent on the velocity-dependent cross-section which carries a subscript). Based on the hierarchical structure formation process, we might expect that the growth of both $R_{\rm vir}$ and $M_{\rm vir}$ should follow some power-law. This is generally supported by our simulations, despite the deviation from the power-law dependence at late times, especially for $R_{\rm vir}$, which may be attributed to the smallness of the simulated box which fails to be statistically representative at low $z$. This can be explored with a larger box size and a greater sample of halos. 

Comparing the three models presented in the figure one can see that the general dependences of $M_{\rm vir}$ and $R_{\rm vir}$ as a function of the scale factor exhibit only minor differences (see Table~\ref{table:halo_evo}). This is expected because the halos considered are high-mass, Milky Way-like halos,  which are not expected to be strongly affected by DM interactions except in their very centers. Indeed, substantial evaporation of massive halos does not occur because the additional random velocities the DM particles acquire in inelastic interactions (i.e., mass conversions), $V_k\sim 100$~km~s$^{-1}$, are substantially smaller than the halo escape velocity.

Another important halo property is the velocity dispersion $\sigma$, which appears to be more or less constant over the time period considered here ($0 \leqslant z \lesssim 3$) with some minor fluctuations. Although this is seen in all the three models, 2cDM and 2cDM$^{\rm conv}_{\rm only}$ show slightly larger fluctuations than SIDM. Interestingly, both of those 2cDM models with a larger DM cross-section show a slight increase in the velocity dispersion compared to a smaller cross-section, which is not seen in the SIDM halos. This effect can be easily understood from the fact that a larger cross-section produces a larger number of inelastic conversions among DM, resulting in an increase in the overall thermal energy within the halo. Thus, `exothermal' conversions $h\to l$ are more probable and dominate over `endothermal' $l\to h$, as is theoretically expected.

It is not surprising to see that the evolution of the halo properties in Figure~\ref{fig:halo_evo} shows minor differences between $\sigma_{0}/m = 0.1$ and 1, contrary to the evolutionary history of the halo core properties shown in Figure~\ref{fig:halocore_evo}. That is because the value of $\sigma_{0}/m$ mostly affects the physics in the halo core, whereas the general, virial halo properties themselves remain largely unaffected over time.

\begin{figure*}
  \centering
  \includegraphics[scale = 0.55]{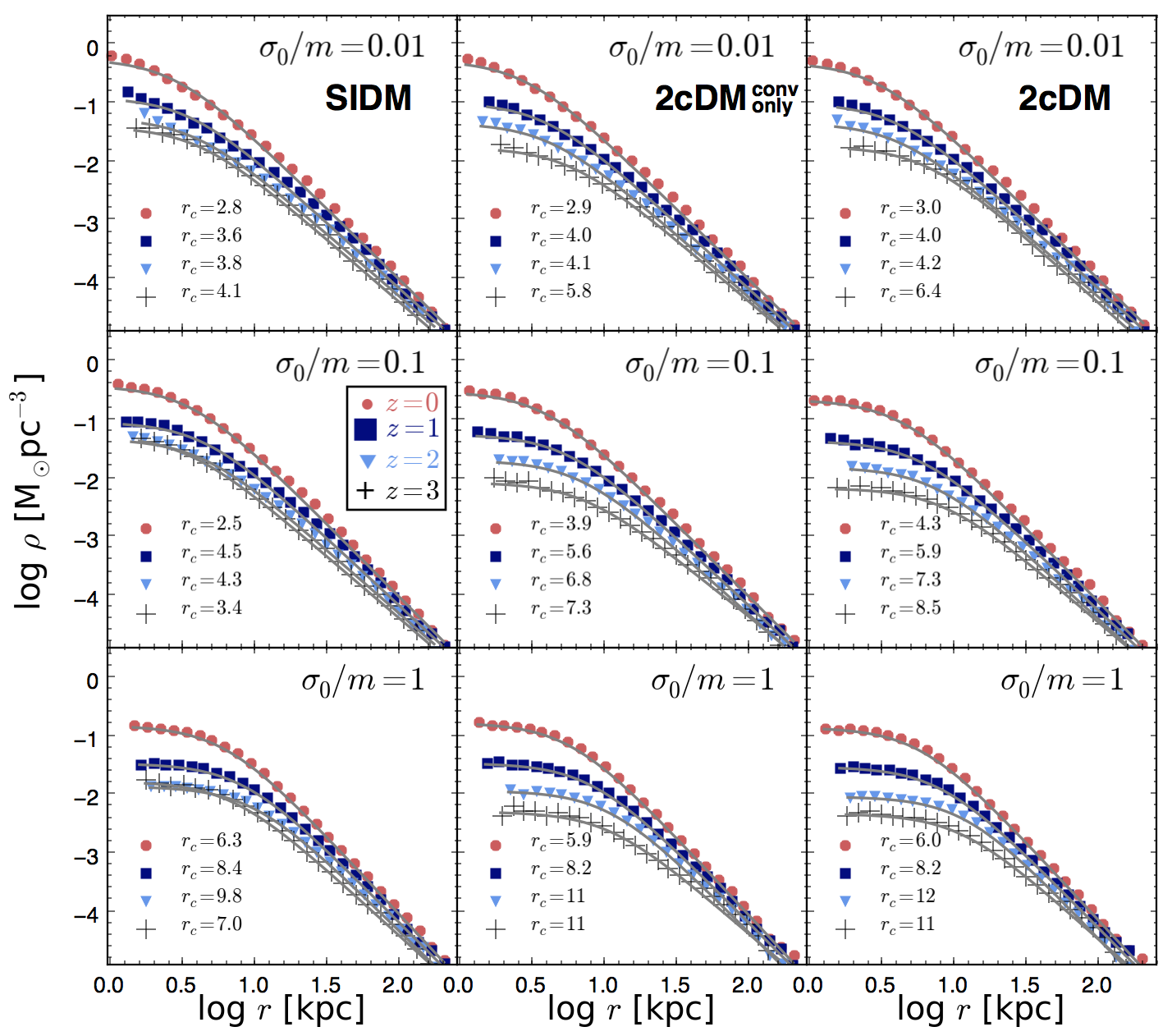}  
 \caption[]{\label{fig:halo-evo-profile} %
Evolution of halo density profiles. The profiles of the largest halo in the simulation are shown. The thin solid line is the gISO fit. For reference, the mean $R_{\rm vir}$/kpc are 208 ($z=0$), 203 ($z=1$), 176 ($z=2$) and 163 ($z=3$). We also show the core radius $r_{c}$ at each redshift in the unit of kpc. The standard deviations are mostly small ($< 1$) with the exception of 2.4 for $z=2$ cases. 
  }
\end{figure*}

\subsection{Halo density profile evolution}
\label{s:prof-evol}

Here we illustrate in more detail the halo profile evolution together with the gISO fits, which has been discussed in previous sections. Figure~\ref{fig:halo-evo-profile} shows the halo density profile evolution at different redshifts ($z=0,1,2$ and 3) for the three models: SIDM, 2cDM$^{\rm conv}_{\rm only}$, and full 2cDM, all based on ($-2,-2$) with $V_{k} = 100$ km s$^{-1}$ as before. Here we only present the profiles of the largest halo in the simulation. We also show the gISO fit for each profile (solid curve) and the core radius $r_{c}$ obtained from it. Based on the fits, it is evident that a core size tends to be larger at earlier times and shrink afterwards, and this appears to be a universal trend among the three models. 
It is also evident that the core densities at earlier times are correspondingly smaller, as discussed earlier. While all the cases indicate larger cores at higher redshifts up until $z=2$, we have some exceptional cases that show slightly smaller cores at $z=3$ compared to that of $z=2$ (bottom row, in particular). This may indicate that the cores have not yet well formed at early redshifts, though numerical resolution may also be an issue. Higher resolution simulations are needed to explore the profiles at high redshifts ($z\gtrsim3$).  

Our results are thus in stark contrast to some of the works that rely on stellar feedback to create cored halos within the framework of the $\Lambda$CDM model. For example, \cite{governato2012} showed that the cored inner profile can be produced \emph{after} the onset of the major star formation process, which occurs around $z=2$. Prior to this burst of star formation, the DM inner profile remains cuspy --- exactly as it is predicted by the traditional $\Lambda$CDM without such stellar feedback. Clearly, future observations should be able to show whether there are any galaxies having cored inner profiles at $z>2$. If there are, it would lend supports to the 2cDM model with ($-2,-2$) in particular. 

A closer look at the profile evolution reveals that the DM mass conversions seem to flatten the inner profile more strongly than the DM elastic scattering \emph{at higher redshift}. In Section \ref{sec:scaling}, we showed that the mass conversions have somewhat a weaker effect on the halo center than the elastic scattering when the results are compared at $z = 0$. However, what we found in the halo evolution history is that the mass conversions begin to affect the halo structure earlier than the elastic scattering. It is also interesting to see that the profiles of SIDM at ($z=2$ and 3) are nearly identical, whereas the other two models with the mass conversions (i.e., 2cDM$^{\rm conv}_{\rm only}$ and 2cDM) show a clear difference in the profiles at $z=2$ and 3. The overall effect of the mass conversions at high $z$ is certainly stronger than that of the elastic scattering alone, as the model that combines the two (2cDM) shows identical profile evolution to a model that only considers the mass conversion (2cDM$^{\rm conv}_{\rm only}$). 

Lastly, we discuss the goodness of the gISO fits to the density profiles. The gISO generally provides very good fits to the cored profiles, which in turn means that it works better with larger DM cross-sections (such as $\sigma_{0}/m = 1$ and 0.1). We see that the smaller cross-section of $0.01$ does produce a core, especially at higher $z$ regardless of the models, but the fit is not as good as for the larger cross-sections with the given numerical resolution. One can visually confirm this by examining the fit's slight deviation from the data points at the inner most part of profiles (top row panels). The reason for this is that the smaller the cross-section, the closer the halo evolution to that in $\Lambda$CDM, and thus the better the profile is fit with NFW. We conclude that the gISO slightly \emph{underestimates} the core density and \emph{overestimates} the core radii in cases with small DM cross-sections.


\section{Velocity profiles} 
\label{sec:Vprofiles}

In addition to density profiles, the velocity profiles also deliver valuable information to probe the inner structure of a halo and are generally easier to compare with observations. Here, we explore the 2cDM parameters with velocity dispersion and circular velocity profiles.

\subsection{Velocity dispersion profiles}

\begin{figure*}
  \centering
  \includegraphics[scale = 0.6]{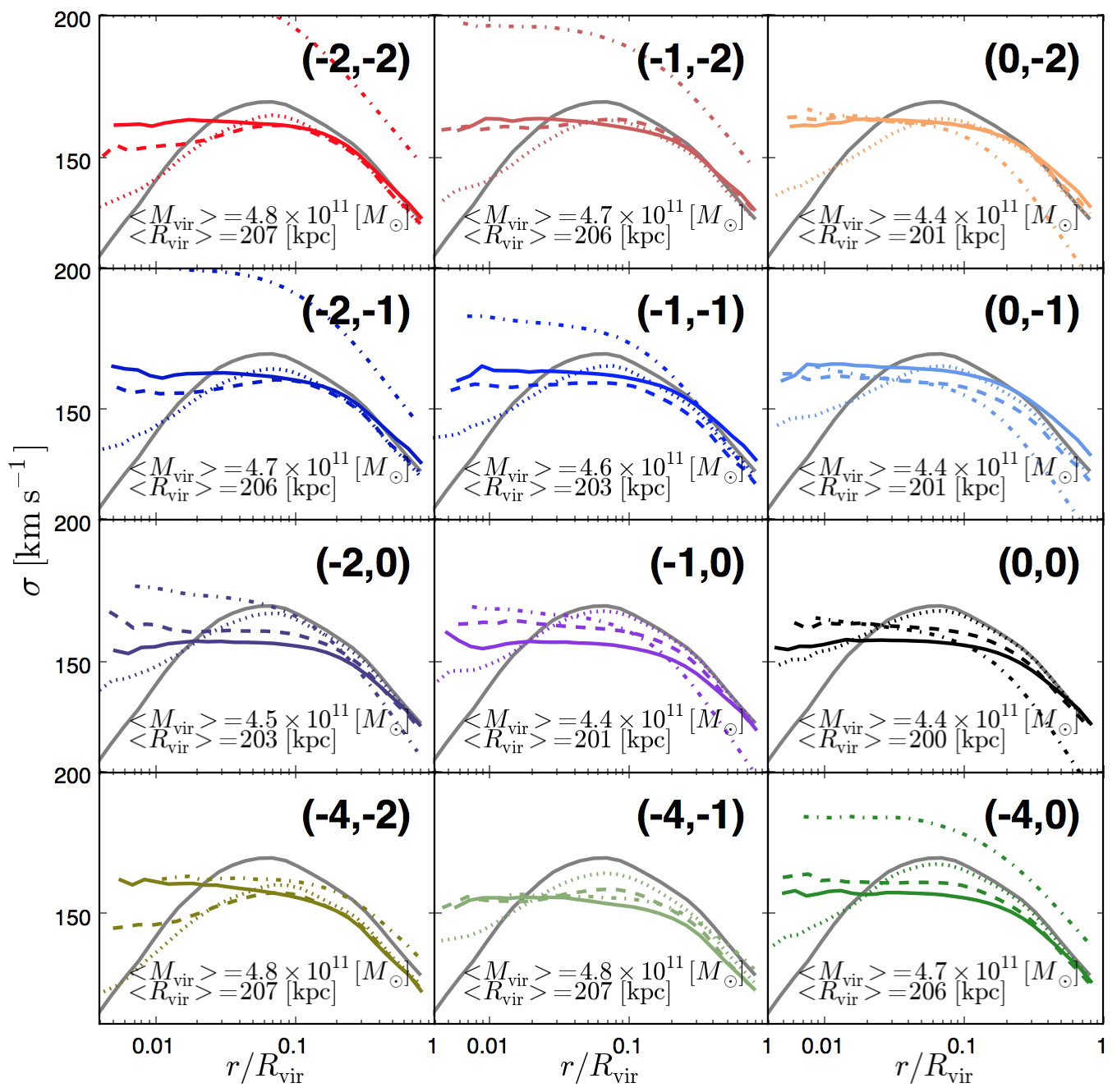}  
  \caption[]{\label{fig:vdisp_all} %
Velocity dispersion profiles. Each panel label represented by a pair of numbers in parentheses represents a specific 2cDM model characterized by the scattering and conversion exponents, $(a_s, a_c)$. Colored curves represent different values of cross-section: the dotted, dashed, solid, and dash-dot lines correspond to $\sigma_{0}/m = 0.01,0.1,1$ and 10, respectively. The thick solid gray line represents CDM, for comparison. 
  }
\end{figure*}

A CDM halo described by the NFW profile has the velocity dispersion decreasing toward the center, i.e., DM distribution is effectively becoming ``cooler'' inward. Elastic scattering of DM particles causes transport of ``heat'' into the center, thus flattening the velocity dispersion profile and making it more isothermal. This consequently reduces the cuspiness of the inner density profile and leads to the formation of a core. In Figure~\ref{fig:vdisp_all}, one can see that the velocity dispersion in the halo center increases as the magnitude of the cross-sections increases in all the 2cDM cases. Even the smallest cross-section used in simulations (i.e. $\sigma_{0}/m = 0.01$) is able to soften the slope of the inner-most profile. This is in stark contrast to the steeply declining inner profile of the CDM model.

However, too large cross-sections are known to lead to the gravothermal catastrophe, associated with the negative heat capacity of self-gravitating systems, see Section \ref{sec:profile_fit} for more discussion. This process results in the core collapse on the time-scale of tens of dynamical times of a core and the formation of an even steeper, $\rho\propto r^{-2}$, cusp typical of isothermal systems. The final inner velocity dispersion is, accordingly, higher than in a non-collapsed halo and approximately flat. This effect is clearly seen in our $\sigma_{0}/m = 10$  cases.


\subsection{Circular velocity profiles}

\begin{figure*}
  \centering
  \includegraphics[scale = 0.6]{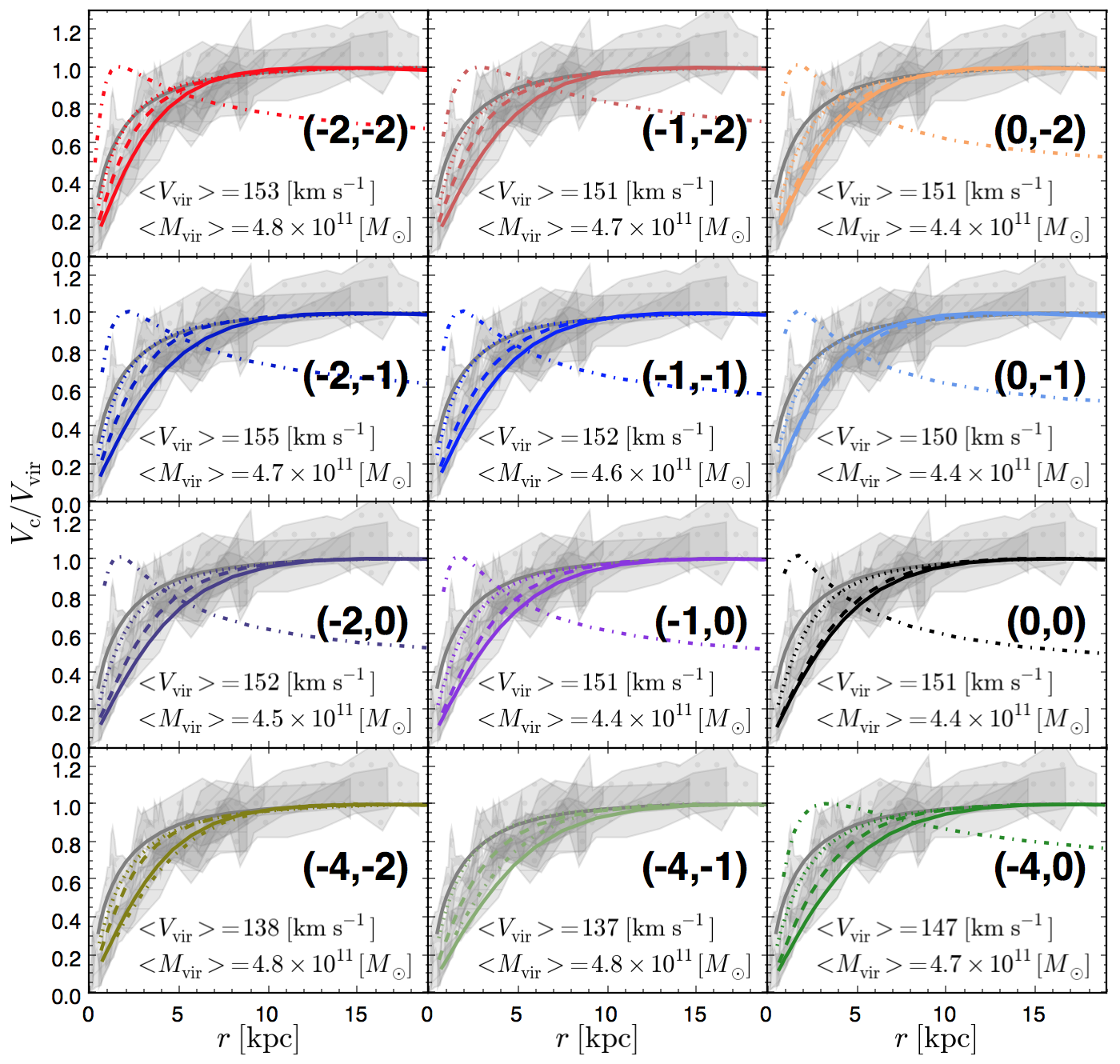}  
  \caption[]{\label{fig:rotcurve_all} %
Rotation curves. The shaded patches are a collection of observed LSB galaxy rotation curves taken from \cite{kuzio2006,kuzio2010} are normalized by the virial velocities. The curve dashing is the same as in Figure \ref{fig:vdisp_all}.
  }
\end{figure*}

Figure~\ref{fig:rotcurve_all} presents the rotation curves for the 2cDM and CDM models (curves) compared with observations (grey shades). 
For better comparison, both the simulated and observed circular velocities were normalized by the virial circular velocities of the halos. 
The dark matter circular velocity profiles are good probes of the dark matter mass distribution. For cases with stronger DM interactions, we would expect the mass concentration in the halo center to be reduced. Indeed, the slope near the halo center ($\lesssim 10$ kpc) is softened for larger cross-sections and deviates from the steeper slope of CDM. It appears that this general trend is common to nearly all the cases with little sensitivity with respect to the values of ($a_{s}, a_{c}$). Comparing with observations, one can see that the most of the 2cDM models and even the CDM are in agreement with observations within the observational uncertainty. 

However, the results of $\sigma_{0}/m = 10$ simulations are at odds with observations. Many of them show a distinctive peak in the mass concentration in the inner-most part of halo (roughly $\sim$2 kpc) followed by a declining profile. Consequently, the $\sigma_{0}/m = 10$ cases, excluding ($-2,-2$), ($-4,-2$) and ($-4,-1$), show a clear discrepancy with observational trends. The formation of the peaky profile indicates the effect of the gravothermal instability, in which the halo core is strongly contracted to form a very steep isothermal cusp. Having the $\sigma_{0}/m = 1$ cases near the lower-end of the observed circular velocity profile, it is likely that cross-sections $\sigma_{0}/m$ greater than a few cm$^2$g$^{-1}$ can be near/past the threshold of the collapse, thereby disagreeing with observations and, thus, should be ruled out.

\section{Conclusion} \label{sec:CN}

In this paper, we have presented the properties of dark matter halos with $M \sim 4-5 \times10^{11} M_{\odot}$ obtained in $N$-body cosmological simulations with the simplest two-component (2cDM) model. A large set of $\sigma_{0}/m$-$ (a_{s},a_{c})$ parameters has been explored. 
Our choice of the parameter set effectively covers four orders of magnitude in the DM cross-section $\sigma_{0}/m$ ranging from 0.01 to 10 cm$^2$ g$^{-1}$, and the DM cross-section's velocity dependences $\sigma \sim v^{a_{s}}$ and $\sim v^{a_{c}}$, for elastic scattering and inelastic mass conversion, respectively, ranging from $a_{s}$ (or $a_{c}$) $= 0$ (i.e. no velocity dependence) to inverse power-law dependence of $a_{s}$ (or $a_{c}$) $= -1$, $-2$, and $-4$ (for $a_{s}$ only). We tested the 2cDM model by performing the total of 57 simulations with medium resolution to throughly explore the combination of the above parameter set. We also examined the time evolution of halo properties from the redshift of $z = 3$ to the present time. 
We summarize our findings as follows.

\emph{Halo density profile}:
We show that 2cDM robustly creates cored inner profiles, thus resolving the core-cusp problem without the need for baryonic feedback.
We find the reduction of the central DM density and the creation of an isothermal core are possible by either elastic scattering or mass conversion alone, or the combination of both. 
That is, both the heat transport towards the halo center by the elastic scattering and mass evaporation by inelastic conversion through DM-DM interactions are sufficient and effective enough to create an isothermal core.
From the comparison of the various set of $\sigma_{0}/m$ and ($a_{s},a_{c}$), we find the magnitude of the DM cross-section is primarily what determines the core size, while the specific choice of ($a_{s},a_{c}$) seems to affect the properties of the halo core less significantly on a Milky Way-like halo. 
In other words, $\sigma_{0}/m = 0.01, 0.1,$ and 1 are all capable of producing a core regardless of the cross-section's velocity dependence. 
The obvious case that can be ruled out based on observations and our results presented in this work is $\sigma_{0}/m = 10$, which tends to produce an steeper inner profile ($\sim r^{-2}$) due to an excessive amount of DM interactions (approaching fluid regime) that is induced by a gravothermal collapse.

\emph{Density profile fit}:
We have presented a generalized isothermal model, gISO, for fitting a cored density profile and quantified the halo core properties.
The model accounts for the flat inner profile and the steep NFW-like outer profile, which gives a surprisingly good fit compared to the pure ISO and NFW models.
Our fit yields a general trend; the larger the core size, the smaller the core density becomes, insensitive to the choice of parameters. 
We consistently see that the core size and the core density are most affected by $\sigma_{0}/m$ rather than the values on ($a_{s},a_{c}$). Although a minor difference, the core size seems to be smaller in general (and correspondingly higher core density) with $a_{s}$ and $a_{c}$ being more negative (for example, $-2$ rather than 0).

\emph{Halo core evolution}:
The gISO fit provides a reliable way to study the time evolution of the halo within the 2cDM paradigm. 
Our study shows the fitting parameters for the halo density profiles, $\rho_{c}$, $r_{c}$ and $\beta$, all generally follow a power-law over the universe scale factor of $0.25 \leq a \leq 1$ ($0 \leq z \leq 3$).
In particular, the evolution of the surface density, $\rho_{c} r_{c}$, shows a very tight power-law relation: $\rho_{c} r_{c} \sim a^{2}$ with $\rho_{c} \sim a^{2.4}$ and $r_{c} \sim a^{-0.4}$ for a particular set of parameters $(a_{s}, a_{c}) = (-2,-2)$ with $\sigma_{0}/m = 0.1$ and 1 in units of cm$^{2}$ g$^{-1}$ and $V_{k} = 100$ km s$^{-1}$.
This universal trend in forming such a tight power-law appears to be common whether the elastic scattering or mass conversion, or both are enabled. We find, however, that the mass conversion does play a more significant role in affecting the halo core properties at higher redshift by producing a larger core size and smaller core density than the case with the elastic scattering alone. 
Despite the effects of the scattering and/or conversion, it is evident that the gradual accretion of mass is still the dominant mechanism to ultimately determine the evolutional trend of the core properties. 
This is implied by the gradual and steady increase in the core density and decrease in the core size over time.
Further, the evolution of $\beta$, which defines the slope of the outer halo in the density profile, suggests that 2cDM halos with ($-2,-2$) tend to have a core that is more isothermal $\beta \sim 2$ at earlier time ($z \sim 3$) than the present time $\beta \sim 2.7$.
The scatter among halo sample is greater, however, but one can still see that $\beta$ follows an approximately power-law relation with the scale factor.

Based on the comparison studies among the available parameter set for 2cDM, primarily on the maximum circular velocity function and DM density profiles, we find that most of the physically motivated choice of parameters on the DM cross-section's velocity dependence and its size are plausible; $-2 \leq a_{s}, a_{c} \leq 0$ and $0.01 \lesssim \sigma_{0}/m \lesssim 1$. Within the framework of 2cDM, $\sigma_{0}/m = 10$ is ruled out partly because it invalidates the rare binary collision approximation as it produces excessive number of DM interactions, which puts it in the fluid regime, and also because it conflicts with the observations by producing a steeper inner density profile owning to gravothermal collapse. 
The case with $a_{s} = -4$ is also generally disfavored for the same reason, although it may still be valid with smaller $\sigma_{0}/m$ values.
Note that our results presented in this work are for halos of $M \lesssim 10^{12} M_{\odot}$ simulated in a rather small box size of $L=3h^{-1}$Mpc, thereby lacking the statistical samples and the environmental effects.

In future work, we will expand our studies by testing the model on dwarfs ($10^7 \sim 10^9 M_{\odot}$) and galaxy clusters ($10^{14} \sim 10^{15} M_{\odot}$) with smaller and larger simulation box sizes. 
Observationally, the halo 'core' size for dwarfs is smaller than that of galaxy clusters, and in fact, the latter appear to have the inner profile only mildly less steeper than that of CDM counterpart \citep{newman2013b}.
It would be interesting to see if 2cDM can reproduce such profiles on both of those systems. 
Our model also needs to be studied with baryonic physics, namely gas and stellar dynamics, to examine the galaxy formation and evolution processes.
There has been ample implications that the presence of baryons and stellar feedback process do affect the DM distributions near the halo centers \citep[e.g.,][]{blumenthal1986,dicintio2014}. 
Moreover, some authors have studied SIDM with baryons in numerical simulations and found that the presence of stellar feedback with SIDM does not significantly alter the results of the CDM counterparts \citep{vogelsberger2014, fry2015}. 
Meanwhile, \citet{kaplinghat2014b} showed SIDM and baryons can be dynamically correlated and affect each other.  
Having shown in this work that the mass conversion in the 2cDM model is the key ingredient in resolving the core-cusp, substructure, and too-big-to-fail problems, which cannot be achieved by elastic scattering or SIDM alone, it is of great interest to study how the additional effects from the presence of baryons and baryonic feedback are intertwined with the 2cDM physics and affect the process of halo and galaxy formation and evolution.


\section*{Acknowledgements}
The simulations for this work have been carried out at the Advanced Computing Facilities at the University of Kansas. 
Special thanks to Jay Neitling for kindly allowing us to use the Open Cluster in the Department of Physics and Astronomy at the University of Nevada, Las Vegas. 
The authors are grateful to Scott Tremaine, Matias Zaldarriaga, Mark Vogelsberger, Lars Hernquist, Ramesh Narayan and Avi Loeb for valuable discussions. MM is grateful to the Institute for Theory and Computation at Harvard University for support and hospitality during his sabbatical leave and acknowledges DOE grant DE-SC0016368.

\bibliographystyle{mnras} 
\bibliography{2cDM-MW-papers12}    

\appendix

\section{Tables of power-law fits of halos at various epochs}

Here we present the tables for the halo fitting parameters as a function of the scale factor. 

\begin{table*}
\centering
\tabcolsep=0.11cm
\begin{tabular}{ccccc}
\hline\hline
& & SIDM & 2cDM$^{\rm conv}_{\rm only}$ & 2cDM \\
\hline

log $\rho_{c}$ [M$_{\odot}$ pc$^{-3}$] & $\sigma_{0} /m = 0.1$, Halo1 & $- 0.44 + 1.9 x$ & $- 0.51 + 2.5 x$ & $- 0.64 + 2.5 x $\\
 & $\sigma_{0} /m = 0.1$, Halo2 & $- 0.67 + 1.7 x $ & $- 0.93 + 2.4 x$ & $- 1.1 + 2.2 x$\\
 & $\sigma_{0} /m = 1$, Halo1 & $- 2.3 + 1.5 x $ & $- 0.76 + 2.4 x$ & $- 0.84 + 2.4 x$\\
 & $\sigma_{0} /m = 1$, Halo2 & $- 2.3 + 1.1 x$ & $- 1.2 + 2.1 x$ & $- 1.3 + 2.3 x $\\
 \hline
log $r_{c}$ [kpc] & $\sigma_{0} /m = 0.1$, Halo1 & $0.59 - 0.16 x$ & $0.61 - 0.48 x$ & $0.64 - 0.50 x $\\
 & $\sigma_{0} /m = 0.1$, Halo2 & $0.56 - 0.021 x$ & $0.67 - 0.24 x$ & $0.71 - 0.29 x$\\
 & $\sigma_{0} /m = 1$, Halo1 & $1.1 -0.26 x$ & $0.79 -0.45 x $ & $0.92 -0.32 x$ \\
 & $\sigma_{0} /m = 1$, Halo2 & $0.83 + 0.10 x$ & $0.92 + 0.11 x$ & $0.79 -0.47 x$\\
 \hline
log $\rho_{c} r_{c}$ [M$_{\odot}$ pc$^{-2}$]& $\sigma_{0} /m = 0.1$, Halo1 & $3.1 + 1.8 x$ & $3.1 + 2.1 x$ & $3.0 + 2.0 x$\\
 & $\sigma_{0} /m = 0.1$, Halo2 & $2.9 + 1.7 x$ & $2.8 + 2.1 x $ &$2.7 + 2.0 x$\\
 & $\sigma_{0} /m = 1$, Halo1 & $1.8 + 1.2 x$ & $3.0 + 2.0 x$ & $3.0 + 1.9 x$ \\
 & $\sigma_{0} /m = 1$, Halo2 & $1.6 + 1.2 x$ & $2.8 + 2.2 x$ & $2.6 + 2.0 x$\\[1ex]
\hline\hline 

\end{tabular}
\caption[]{ Fitting parameters regarding the halo core properties with respect to the scale factor. Here we define $x \equiv \log a$, where $a$ is the scale factor. $\rho_{c}, r_{c}$, $\rho_{c} r_{c}$ and $\sigma_{0}/m$ are the core density, core radius, surface density, and the DM cross section per unit mass, respectively.  The fit was conducted on the two largest, most resolved halos ('Halo1' and 'Halo2') in the simulations.
 }
\label{table:halocore_evo}
\end{table*}

\begin{table*}
\centering
\tabcolsep=0.11cm
\begin{tabular}{ccccc}
\hline\hline
& & SIDM & 2cDM$^{\rm conv}_{\rm only}$ & 2cDM \\
\hline

log $M_{\rm vir}$[M$_{\odot}$] & $\sigma_{0} /m = 0.1$, Halo1 & $12 + 0.92 x$ & $12 + 0.94 x$ & $12 + 0.94 x$\\
 & $\sigma_{0} /m = 0.1$, Halo2 & $12 + 1.1 x$ & $12 + 1.2 x$ & $11 + 1.1 x $\\
 & $\sigma_{0} /m = 1$, Halo1 & $11 + 0.52 x$ & $12 + 0.92 x$ & $12 + 0.94 x$\\
 & $\sigma_{0} /m = 1$, Halo2 & $11 + 0.76 x$ & $12 + 1.2 x $ & $11 + 1.0 x $\\
 \hline
log $R_{\rm vir}$ [kpc] & $\sigma_{0} /m = 0.1$, Halo1 & $2.4 + 0.19 x$ & $2.3 + 0.21 x$ & $2.4 + 0.21 x$\\
 & $\sigma_{0} /m = 0.1$, Halo2 & $2.3 + 0.23 x$ & $2.3 + 0.30 x$ &$2.3 + 0.24 x$\\
 & $\sigma_{0} /m = 1$, Halo1 & $2.3 + 0.079 x$ & $2.3 + 0.20 x $ & $2.3 + 0.21 x$ \\
 & $\sigma_{0} /m = 1$, Halo2 & $2.1 + 0.16 x$ & $2.3 + 0.29 x$ & $2.3 + 0.22 x$\\
 \hline
log $\sigma$ [km s$^{-1}$]& $\sigma_{0} /m = 0.1$, Halo1 & $2.1 -0.040 x$ & $2.1 -0.039 x$ & $2.1 -0.040 x$ \\
 & $\sigma_{0} /m = 0.1$, Halo2 & $2.1 -0.041 x$ & $2.0 -0.012 x$ & $2.0 -0.029 x$\\
 & $\sigma_{0} /m = 1$, Halo1 & $2.2 -0.055 x$ & $2.1 -0.038 x$ & $2.1 -0.027 x $ \\
 & $\sigma_{0} /m = 1$, Halo2 & $2.1 -0.034 x$ & $2.0 + 0.020 x$ & $2.0 -0.016 x $ \\[1ex]
\hline\hline 

\end{tabular}
\caption[]{ Fitting parameters regarding the virial halo properties with respect to the scale factor. 
The virial mass ($M_{\rm vir}$), virial radius ($R_{\rm vir}$), and the velocity dispersion ($\sigma$) of the halos are computed and extracted by the Amiga Halo Finder \citep{knollmann2009}. The other symbols are defined the same as in Table~\ref{table:halocore_evo}.
}
\label{table:halo_evo}
\end{table*}

\section{Convergence test}
\begin{figure}
  \centering
  \includegraphics[scale = 0.4]{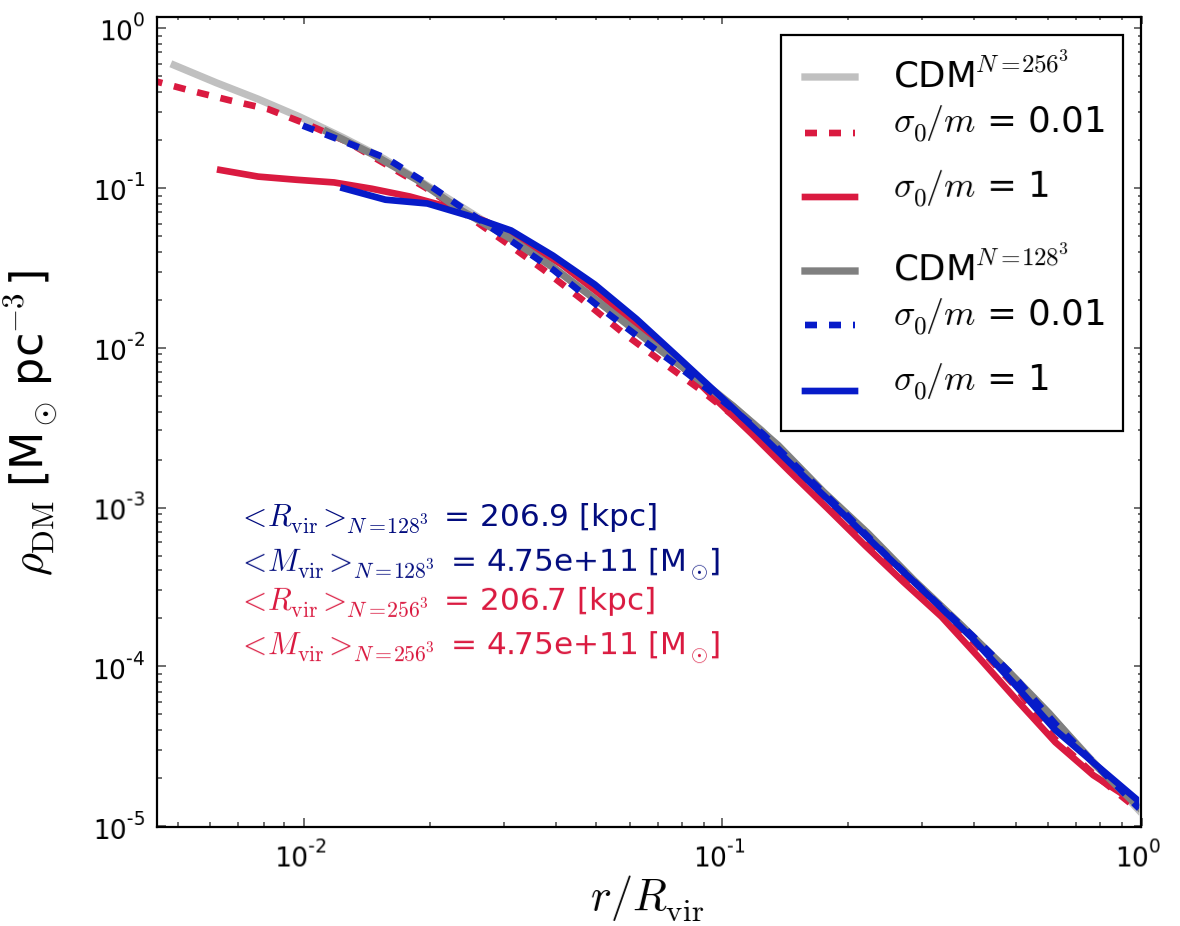}  
  \caption[]{\label{fig:convergence_profiles} %
Convergence test -- Halo density profile: $N=128^3$ (blue) and $N=256^3$ (red). 2cDM with two different cross-section sizes of $\sigma_{0}/m = 0.01$ and 1 are being compared with CDM. The profiles of the largest, most well-resolved halo, are shown. The virial radius and mass of the halo in the two resolutions are virtually identical.
  }
\end{figure}

We compared two cases with the total number of particles in the simulation box of $N = 256^{3}$ and $128^{3}$ to check our model's dependency on numerical resolution. Figure~\ref{fig:convergence_profiles} shows a good convergence overall on a well-resolved halo. That is, the cases with $N = 256^{3}$ appears to be just the extensions of the $N = 128^{3}$ cases without significantly altering the halo profiles.
The $\sigma_{0}/m =0.01$ and 1 do not seem to be affected by the resolution difference.

\label{lastpage}

\end{document}